\documentstyle[twoside,psfig,epsfig]{article}

\catcode`\@=11
\long\def\@makefntext#1{
\protect\noindent \hbox to 3.2pt {\hskip-.9pt
$^{{\eightrm\@thefnmark}}$\hfil}#1\hfill}               %CAN BE USED

\def\@makefnmark{\hbox to 0pt{$^{\@thefnmark}$\hss}}    %ORIGINAL

\def\ps@myheadings{\let\@mkboth\@gobbletwo
\def\@oddhead{\hbox{}
\rightmark\hfil\eightrm\thepage}
\def\@oddfoot{}\def\@evenhead{\eightrm\thepage\hfil
\leftmark\hbox{}}\def\@evenfoot{}
\def\sectionmark##1{}\def\subsectionmark##1{}}
\oddsidemargin=\evensidemargin
\newcounter{sectionc}\newcounter{subsectionc}\newcounter{subsubsectionc}
\renewcommand{\section}[1] {\vspace{12pt}\addtocounter{sectionc}{1}
\setcounter{subsectionc}{0}\setcounter{subsubsectionc}{0}\noindent
        {\tenbf\thesectionc. #1}\par\vspace{5pt}}
\renewcommand{\subsection}[1] {\vspace{12pt}\addtocounter{subsectionc}{1}
        \setcounter{subsubsectionc}{0}\noindent
        {\bf\thesectionc.\thesubsectionc. {\kern1pt \bfit #1}}\par\vspace{5pt}}
\renewcommand{\subsubsection}[1] {\vspace{12pt}\addtocounter{subsubsectionc}{1}
        \noindent{\tenrm\thesectionc.\thesubsectionc.\thesubsubsectionc. 
        {\kern1pt \tenit #1}}\par\vspace{5pt}}

\newcounter{appendixc}
\newcounter{subappendixc}[appendixc]
\newcounter{subsubappendixc}[subappendixc]
\renewcommand{\thesubappendixc}{\Alph{appendixc}.\arabic{subappendixc}}
\renewcommand{\thesubsubappendixc}
        {\Alph{appendixc}.\arabic{subappendixc}.\arabic{subsubappendixc}}

\renewcommand{\appendix}[1] {\vspace{12pt}
        \refstepcounter{appendixc}
        \setcounter{figure}{0}
        \setcounter{table}{0}
        \setcounter{lemma}{0}
        \setcounter{theorem}{0}
        \setcounter{corollary}{0}
        \setcounter{definition}{0}   
        \setcounter{equation}{0}
        \renewcommand{\thefigure}{\Alph{appendixc}.\arabic{figure}}
        \renewcommand{\thetable}{\Alph{appendixc}.\arabic{table}}
        \renewcommand{\theappendixc}{\Alph{appendixc}}
        \renewcommand{\thelemma}{\Alph{appendixc}.\arabic{lemma}}
        \renewcommand{\thetheorem}{\Alph{appendixc}.\arabic{theorem}}
         \renewcommand{\thedefinition}{\Alph{appendixc}.\arabic{definition}}
        \renewcommand{\thecorollary}{\Alph{appendixc}.\arabic{corollary}}
        \renewcommand{\theequation}{\Alph{appendixc}.\arabic{equation}}
        \noindent{\tenbf Appendix \theappendixc #1}\par\vspace{5pt}}
\newcommand{\subappendix}[1] {\vspace{12pt}
        \refstepcounter{subappendixc}
        \noindent{\bf Appendix \thesubappendixc. {\kern1pt \bfit #1}}
        \par\vspace{5pt}}
\newcommand{\subsubappendix}[1] {\vspace{12pt}
        \refstepcounter{subsubappendixc}
        \noindent{\rm Appendix \thesubsubappendixc. {\kern1pt \tenit #1}}
        \par\vspace{5pt}}

\newcommand{\textlineskip}{\baselineskip=13pt}
\newcommand{\smalllineskip}{\baselineskip=10pt}

\def\eightcirc{
\begin{picture}(0,0)
\put(4.4,1.8){\circle{6.5}}
\end{picture}}
\def\eightcopyright{\eightcirc\kern2.7pt\hbox{\eightrm c}}

\renewenvironment{thebibliography}[1]
        {\frenchspacing
         \ninerm\baselineskip=11pt
         \begin{list}{\arabic{enumi}.}
        {\usecounter{enumi}\setlength{\parsep}{0pt}
         \setlength{\leftmargin 12.7pt}{\rightmargin 0pt} %FOR 1--9 ITEMS
         \setlength{\itemsep}{0pt} \settowidth
        {\labelwidth}{#1.}\sloppy}}{\end{list}}

\newcounter{itemlistc}
\newcounter{romanlistc}
\newcounter{alphlistc}
\newcounter{arabiclistc}

\newcommand{\fcaption}[1]{
        \refstepcounter{figure}
        \setbox\@tempboxa =
\hbox{\footnotesize Fig.~\thefigure. #1}
        \ifdim \wd\@tempboxa > 5in
           {\begin{center}
           \parbox{5in}{\footnotesize\smalllineskip Fig.~\thefigure. #1}
            \end{center}}
        \else
             {\begin{center}
             {\footnotesize
Fig.~\thefigure. #1}
              \end{center}}
        \fi}

\newcommand{\tcaption}[1]{
        \refstepcounter{table}
        \setbox\@tempboxa = \hbox{\footnotesize Table~\thetable. #1}
        \ifdim \wd\@tempboxa > 5in
           {\begin{center}
        \parbox{5in}{\footnotesize\smalllineskip Table~\thetable. #1}
            \end{center}}
        \else
             {\begin{center}
             {\footnotesize Table~\thetable. #1}
              \end{center}}
        \fi}

\def\@citex[#1]#2{\if@filesw\immediate\write\@auxout
        {\string\citation{#2}}\fi
\def\@citea{}\@cite{\@for\@citeb:=#2\do
        {\@citea\def\@citea{,}\@ifundefined
        {b@\@citeb}{{\bf ?}\@warning
        {Citation `\@citeb' on page \thepage \space undefined}}
        {\csname b@\@citeb\endcsname}}}{#1}}

\newif\if@cghi
\def\cite{\@cghitrue\@ifnextchar [{\@tempswatrue
        \@citex}{\@tempswafalse\@citex[]}}
\def\citelow{\@cghifalse\@ifnextchar [{\@tempswatrue
        \@citex}{\@tempswafalse\@citex[]}}
\def\@cite#1#2{{$\null^{#1}$\if@tempswa\typeout
        {IJCGA warning: optional citation argument
        ignored: `#2'} \fi}}

\def\pmb#1{\setbox0=\hbox{#1}
        \kern-.025em\copy0\kern-\wd0
        \kern.05em\copy0\kern-\wd0
        \kern-.025em\raise.0433em\box0}

\def\fnt#1#2{\footnotetext{\kern-.3em
        {$^{\mbox{\scriptsize #1}}$}{#2}}}
        
\def\fpage#1{\begingroup
\voffset=.3in
\thispagestyle{empty}\begin{table}[b]\centerline{\footnotesize #1}
        \end{table}\endgroup}
              
\font\tenrm=cmr10
\font\tenit=cmti10
\font\tenbf=cmbx10
\font\bfit=cmbxti10 at 10pt
\font\ninerm=cmr9

\font\eightrm=cmr8

\def\be{\begin{equation}}
\def\ee{\end{equation}}
\def\simlt{\stackrel{<}{{}_\sim}}
\def\simgt{\stackrel{>}{{}_\sim}}
\def\qed{\hbox{${\vcenter{\vbox{                        %HOLLOW SQUARE   
   \hrule height 0.4pt\hbox{\vrule width 0.4pt height 6pt
   \kern5pt\vrule width 0.4pt}\hrule height 0.4pt}}}$}}

\begin{document}
\setlength{\textheight}{7.7truein}  %for 2nd page onwards

%\runninghead{The Higgs Sector and Electron Electric Dipole Moment $\ldots$}
%{M. Boz}

\normalsize\textlineskip
\thispagestyle{empty}
\setcounter{page}{1}

%\copyrightheading{}                     %{Vol. 0, No.0 (1992) 000--000}

%\vspace*{0.88truein} 

\fpage{1}
\centerline{\bf THE HIGGS SECTOR AND ELECTRON ELECTRIC DIPOLE MOMENT}
\centerline{\bf IN NEXT--TO--MINIMAL SUPERSYMMETRY}
\centerline{\bf WITH EXPLICIT CP VIOLATION}
\vspace*{0.37truein}
\centerline{\footnotesize M\"{U}GE BOZ}
\baselineskip=12pt
\centerline{\footnotesize\it Physics  Department, Hacettepe University}
\baselineskip=10pt
\centerline{\footnotesize\it Ankara, 06532, Turkey }
\vspace*{10pt}
%%%%%%%%%%%%%%%%%%%%%%%%%%%%%%%%%%%%%%%%%%%%%%%%%%%%%%%%%%%%%%%%%%%%%%%%%
\begin{abstract}
\noindent
We study the explicit CP violation of the Higgs sector in the 
next--to--minimal supersymmetric model with a gauge singlet Higgs field.
Our numerical predictions show that electric dipole moment 
of electron lies around the present experimental upper 
limits. The mass of the lightest Higgs boson
is quite sensitive to 
the CP violating phases in the
theory. 
It is observed that as the vacuum expectation value of 
the singlet gets higher values,  CP violation increases.\\
%PACS:  12.60.Jv, 11.30.Er, 13.40.Em\\
%Key-Words: supersymmetry, CP violation, electron electric dipole moment
\end{abstract}
%%%%%%%%%%%%%%%%%%%%%%%%%%%%%%%%%%%%%%%%%%%%%%%%%%%%%%%%%%%%%%%%%%%%%%%%%%%%%%%%%%%%%%%%%%%%%%%%%%%%%%%%%%
\section{Introduction}
The present bounds on the electric dipole moments (EDM) of the particles
generate serious hierarchy problems concerning the amount of CP violation. 
For instance,  the neutron EDM~\cite{Harris}, induced by the  the QCD
vacuum angle ($\theta_{QCD}$), is approximately ten orders of magnitude larger
than the existing bound, which is the source of the so-called strong CP
problem. This naturalness problem has been  solved by the Peccei--Quinn
mechanism~\cite{Peccei} 
which promotes $\theta_{QCD}$  to a dynamical variable via the phases
of the quarks or additional color triplets~\cite{Peccei,Weinberg,Kim}.

In the supersymmetric (SUSY) extensions of the standard electroweak theory (SM) 
this hierarchy problem still persists. Moreover, there appear novel sources of 
CP violation coming from the soft supersymmetry breaking mass terms. Though 
the phases of the soft terms have been shown to relax to CP--conserving points
in the minimal model (MSSM)~\cite{Dimopoulos}, this is not necessarily true in the 
non--minimal model (NMSSM)~\cite{Demir1} containing a singlet. 
The Lagrangian of the MSSM
consists of various mass parameters which are not necassarily real~\cite{Dugan}.
The phases of these parameters contribute to known  CP violation observables
such as the electric dipole moments
of the electron, neutron, atoms, and
molecules~\cite{Ibrahim,Ritz1,Abel1,Pilaftsis3,Lebedev1,BozKaragoz}, the Higgs
system~\cite{Pilaftsis1,Boz1,BozPak}, 
and the decays and mixings of mesons~\cite{DemirOlive}. 

In addition to these CP hierarchy problems, in minimal SUSY 
model there is another hierarchy problem concerning the Higgsino Dirac 
mass parameter ($\mu$), that is, this mass parameter follows from the 
superpotential of the model and there is no telling of at what 
scale (ranging from $M_W$ to $M_{Pl}$) it is stabilized.

The next--to--minimal supersymmetric
model (NMSSM)
is the most economic extension of the MSSM in which the
$\mu$ parameter is induced by the vacuum expectation value (VEV) of 
an additional gauge singlet. 
The NMSSM not only solves the $\mu$ problem
by means of the VEV of the singlet,
but also offers a rich
phenomenology for colliders~\cite{Demir3}.

In the NMSSM, the Higgs sector  
of the MSSM (including two CP-even Higgs bosons,  a CP odd Higgs boson, and a charged Higgs 
pair) is extended  so that there are three CP-even Higgs bosons, two CP-odd Higgs 
bosons, and a charged Higgs pair. Therefore, the Higgs phenomenology 
of the NMSSM significantly differ from that of MSSM~\cite{Ellwanger2005,Miller2004}.
As a result of the SM  searches at LEP, the lower bound on the lightest Higgs boson mass 
is 115~$\mbox{GeV}$~(and correspondingly $\tan\beta \simgt 3.5$)~\cite{LEP2}. Therefore,
from the searches at LEP2, the lower limit on the mass of the SM Higgs boson
excludes the substantial part of the MSSM parameter space particularly 
(for $m_t=175~\mbox{GeV}$)
at low $\tan\beta$ ($\tan\beta \simlt 3.5$). 
However, in the low $\tan\beta$  regime, NMSSM is in much better shape phenomenologically, since 
the Higgs boson masses are larger, and the 
fine-tuning is less~\cite{Bastero-Gil2000,Dermisek2005}.
Thus, the 
phenomenological implications of the NMSSM 
can offer new opportunities 
at future  colliders~\cite{Ellwanger2005}.

In the MSSM, since the tree level
vacua are CP conserving, spontaneous CP violation (SCPV) can only occur if the
radiative corrections are taken into account leading to a very light
Higgs boson which has been discarded by the experiment~\cite{Maekawa}.
In the NMSSM, 
spontaneous CP violation can occur even at the tree level,
for those models without a discrete $Z_3$ symmetry~\cite{Branco,Davies}.
For models which has $Z_3$ symmetry,
the NMSSM can not produce SCPV at the tree level~\cite{Romao1986}.
SCPV can only occur if the stop quark masses at the one-loop effective potential 
are non-degenerate~\cite{nmssm4}. However, at the tree level,   
unlike the MSSM, explicit CP violation is possible in the NMSSM~\cite{Haba,nmssm2}.

In the  literature, 
the NMSSM has been studied through many researches
including the effects of one-loop  radiative corrections 
due to various particles, and their
superpartners~\cite{nmssm0,HamKimOhSon1,Ham2,HamKimOhSon2,Panagiotakopoulos},
and  the analysis of the Higgs potential with arbitrary number of Higgs
singlets~\cite{Ham5}. In the earlier studies,  
the corrections to mass matrices of the top and stop quarks 
have been calculated 
with either the one-loop effective potential or the 
renormalization group approach~\cite{nmssm0}.
Among recent  works the authors of Ref.~[32] 
particularly addressed the effects of the explicit CP violation 
with the emphasis on the charged Higgs boson in the NMSSM, using the effective potential method,
for which case the CP violating  phases are induced from the 
stop and sbottom quark masses. This analysis is extended 
by including the contribution of the charginos~\cite{Ham2}, and 
of the  neutralinos~\cite{HamKimOhSon2}  to the one-loop effective potential, 
where the tree level CP violating phase is chosen to be equal to the 
one at the loop-level, and the 
complex phases of the chargino and 
neutralino contributions in the radiative corrections are taken into 
consideration.

The radiatively induced CP violation effects 
in the NMSSM has also been analyzed with 
renormalization group improvement~\cite{ellis1}.
It has been shown in Ref.~[37] that 
the renormalization group analysis 
of  radiative symmetry breaking always leads to $\tan\beta$
values that are larger than 1. Indeed this result 
is a general feature of all phenomenologically viable theries 
that incorporate low energy supersymmetry and radiative 
breaking of the electroweak supersymmetry. 
More recent analysis on the Higgs mass spectrum 
motivated by the renormalization group analysis shows  that
to prevent the mass splitting between the light and heavy Higgs bosons  from 
becoming too large, the value of $\tan\beta$  
should be kept moderate ($\simlt 10$)~\cite{Miller}, which we shall also assume in this work.

In this work our aim is to investigate the CP violation capability of 
the NMSSM. Therefore, we limit  our analysis to the effective potential 
with no renormalization group improvement.
This accuracy 
has proven sufficient in obtaining the observable 
effects of explicit CP violation 
on the Higgs masses~\cite{HamKimOhSon1}. In doing this, 
we will keep the  value of $\tan\beta$  
moderate ($\simlt 10$), 
as favoured by the renormalization group analysis of 
[38], which is a well-motivated parameter regime 
for the model under concern
(e.g, $\tan\beta=1.5$ is allowed within the 
framework of NMSSM~\cite{ellis1}).  
For the purpose of showing the 
mass spectra, we shall take   $\tan\beta=2$. To give feeling of the 
sensitivity of the mass on $\tan\beta$, we shall also consider 
$\tan\beta=10$ regime.

In the following we will study the 
radiatively corrected Higgs masses and mixings,
taking into account of the CP violation effects, in the parameter 
space allowed by the electron electric dipole moment (eEDM) constraint.  
We will base our calculations to those of Ref.~[32]. 
The main difference with the 
previous work~\cite{HamKimOhSon1}  springs from the fact that 
in our analysis 
the CP violating phases at the tree level and at the one-loop 
are not equal. Therefore,
we analyze the mass spectra at the tree and one-loop levels, 
as well as the  CP odd components of the Higgs boson 
by  considering the effects of physical   phases seperately.
In the numerical analysis, we particularly 
focus on the  real and complex cases of the tree-level coupling  $\lambda$.
Analyzing these specific cases 
for various parameter planes,
we  search for the  impacts of the tree and 
one-loop level contributions  on the  Higgs sector
which give the opportunity of comparing the results.
In our analysis, we focus on the regions of the  parameter space 
which are allowed by  the  eEDM constraint.

The organization of the paper is as follows:
In Section 2, we study the 
Higgs masses and the one-loop eEDM  in the 
NMSSM. In Section 3,  we discuss two special cases of $\lambda$ 
for two different values of $v_s$, and $A_t$ at $\tan\beta=2$,
and $\tan\beta=10$  regimes. In Section 4, we conclude the work.

\section{The Model}
The Higgs potential of the model at the tree level 
is given by :
\begin{eqnarray}
V_0&=&m_{H_{d}}^2 |H_d|^2+m_{H_{u}}^2 |H_u|^2+
m_S^2 |S|^2- \big [ \lambda A_\lambda H_d H_u S+ \frac{k}{3} A_k S^3+ h.c \big]
\nonumber\\
&+& |\lambda|^2 \big[ \big (|H_d|^2 +  |H_u|^2 \big)|S|^2+ |H_d H_u|^2 \big]+|k|^2 |S|^4 
-(\lambda k^{*} H_d H_u {S^{*}}^2+ h.c)\nonumber\\ 
&+&\frac{1}{8}\big(g_1^2+g_2^2\big) \big(|H_d|^2 - |H_u|^2\big)^2~,
\label{softMSSM}
\end{eqnarray}
where the first, second and the third lines represent
the soft SUSY breaking terms, F and D term contributions, respectively.
Here, $A_\lambda$, and $A_k$ are the for the trilinear soft SUSY breaking terms.

Assuming  $\lambda A_\lambda$, and  $k  A_k$
are real and positive,  the phase  between $\lambda$  and  $k^*$   
is given by:
\begin{eqnarray}
\varphi_{k \lambda}&=&Arg[\lambda k^*]=\varphi_{\lambda}-\varphi_{k}~,
\end{eqnarray}
and this phase forms the source of CP violation at the tree level in Eq. (1).

After electroweak breaking, the Higgs doublets and one Higgs singlet  
in (\ref{softMSSM}) can be expanded as: 
\begin{eqnarray}
\label{doublet}
H_{d}&=&\left(\begin{array}{c c} H_{d}^{0}\\
H_{d}^{-}\end{array}\right)
=\left(\begin{array}{c c}
v_{d}+\phi_{1}+i\varphi_{1}\\ H_{d}^{-}\end{array}\right)\;,\nonumber\\
H_{u}&=&\left(\begin{array}{c c} H_{u}^{+}\\
H_{u}^{0}\end{array}\right)
=\left(\begin{array}{c c}
H_{u}^{+}\\ v_{u}+\phi_{2}+i\varphi_{2}\end{array}\right)~, 
\end{eqnarray}
and 
\begin{eqnarray}
S&=&v_s+\phi_s+i\varphi_s~.
\end{eqnarray}

As usual, we calculate the Higgs masses and their mixings up to one loop accuracy via
\begin{eqnarray}
M^{2}=\left(\frac{\partial^{2}\ V} {\partial \chi_{i} 
\partial \chi_{j}}\right)_{0},~\mbox{where}\; 
\chi_{i} \in {\cal{B}}=\{ \phi_{1}, \phi_{2}, \varphi_{1}, \varphi_{2},
\phi_s, \varphi_s \}~. 
\end{eqnarray}
Here, $V \equiv V_0 + V_{1-loop}$ is the radiatively corrected Higgs
potential~\cite{Ham2}, and as we 
mentioned before we take into account the 
top-stop and bottom-sbottom loop corrections.

The stop and sbottom mass-squared eigenvalues are given by: 
\begin{eqnarray}
\label{stopx}
m_{\tilde{t}_{1,2}}^{2}&=&m_t^2+\frac{1}{2}(M_{Q}^2+M_{T}^2) \pm
\Delta_{\tilde{t}}^{2}~,\nonumber\\
m_{\tilde{b}_{1,2}}^{2}&=&m_b^2+\frac{1}{2}(M_{Q}^2+M_{T}^2) \pm  \Delta_{\tilde{b}}^{2}~,
\end{eqnarray}
where the stop and sbottom mass-splittings read as: 
\begin{eqnarray}
\label{sbotx}
\Delta_{\tilde{t}}^{2}&=&\sqrt{ \frac{1}{4}( M_{Q}^2-M_T^2)^2
+m_{t}^{2} \Big(A_{t}^{2}+\lambda^2 v_s^2 ( t^{-1}_{\beta})^{2}+2 A_{t}
\lambda  v_s  t^{-1}_{\beta}  \cos\varphi_{\lambda t}\Big)}~,\nonumber\\
\label{deltat}
\Delta_{\tilde{b}}^{2}&=&\sqrt{  \frac{1}{4}( M_{Q}^2-M_T^2)^2
+m_{b}^{2} \Big(A_{b}^{2}+\lambda^2 v_s^2 ( t_{\beta})^{2}+2 A_{b}
\lambda   v_s  t_{\beta} \cos\varphi_{\lambda t} \Big)}~.
\end{eqnarray}
Here,
$t_{\beta}$=$\tan\beta$,\,\, $t^{-1}_{\beta}$=$\cot\beta$.
For  convenience, we set 
the soft SUSY breaking scalar-quark
masses as $M_{\tilde{Q}} =M_{\tilde{u}}=M_{\tilde{d}}$,
and  the squark trilinear couplings as $A_{t}=A_{b}$.

The stop and sbottom mass splittings depend explicitely 
on the total CP violation angle
$\varphi_{\lambda t}$ between  $A_t=A_b$, and $\lambda$  
\begin{eqnarray}
\varphi_{\lambda t}&=&Arg[\lambda A_t]~,
\end{eqnarray}
which forms the source of CP violation at the tree and one-loop level Higgs
potential of the NMSSM.

The  (5$\times$5) dimensional Higgs mass--squared matrix  can be expressed as: 
\begin{eqnarray}
M_{ij}=M_{ij}+\Delta M_ {ij}~.
\label{pot}
\end{eqnarray}
Here, $M_{ij}$ comes from the tree-level potential, whereas 
$\Delta M_ {ij}$ from the stop and sbottom contributions at the one-loop
level~\cite{HamKimOhSon1}.

The  elements of the mass matrix at the tree level are  are given by:
\begin{eqnarray}
M_{11}&=&\big[m_Z \cos \beta\big]^2+ \big [m_A \sin \beta\big]^2~, \nonumber\\ 
M_{12}&=& -\big[ m_Z^2+ m_A^2-2 \lambda^2 v^2\big] \sin\beta \cos\beta~, \nonumber\\
M_{13}&=&0~,\nonumber\\
M_{14}&=&-\bigg( \frac{v}{v_s}\bigg ) \bigg [ m_A^2 \sin^2 \beta \cos\beta  -2 
(\lambda^2 v_s^2) \cos \beta + \bigg (\frac{k}{\lambda}\bigg ) (\lambda^2 v_s^2)  \cos \varphi_{\lambda t}    \sin
\beta  \bigg],\nonumber\\
M_{15}&=&-3 \bigg(\frac{v}{v_s}\bigg)\bigg[ \bigg ( \frac {k} {\lambda }\bigg) (\lambda^2 v_s^2)
\sin \beta \sin \varphi_{\lambda t }\bigg]~,\nonumber\\
M_{22}&=&\big [m_Z \sin \beta\big]^2+ \big[m_A \cos \beta\big]^2~, \nonumber\\ 
M_{23}&=& 0~,\nonumber\\
M_{24}&=&-\bigg (\frac{v}{v_s}\bigg ) \bigg [ m_A^2 \sin\beta \cos^2 \beta  -2 (\lambda^2 v_s^2)
\sin \beta   + \bigg (\frac {k} {\lambda }\bigg) (\lambda^2  v_s^2) \cos\varphi_{k \lambda} \cos
\beta  \bigg]~,\nonumber\\
M_{25}&=&M_{52}=-3  \bigg(\frac{v}{v_s}\bigg) \bigg [  \bigg (\frac {k} {\lambda }\bigg)  (\lambda^2 v_s^2) \sin
\varphi_{\lambda t} \cos \beta \bigg]~,\nonumber\\
M_{33}&=& m_A^2~,\nonumber\\
M_{34}&=& \bigg(\frac{v}{v_s}\bigg) \bigg[ \bigg ( \frac {k} {\lambda }\bigg)  (\lambda^2 v_s^2)  \sin \varphi_{
\lambda t}\bigg]~,\nonumber\\
M_{35}&=&\bigg(\frac{v}{v_s}\bigg) \bigg [ m_A^2 \sin \beta \cos \beta -3 \bigg(\frac{k}{\lambda}\bigg)
(\lambda^2 v_s^2) \cos \varphi_{ \lambda t } \bigg]~, \nonumber\\
M_{44}&=&  \bigg( \frac{v^2}{v_s^2}\bigg)\sin\beta \cos\beta \bigg(\sin\beta \cos\beta m_A^2-
\bigg (\frac{k}{\lambda}\bigg)  (\lambda^2 v_s^2) \cos \varphi_{ \lambda t} \bigg]\nonumber\\
&+& \bigg (\frac{k^2} {\lambda^2}\bigg)(\lambda^2 v_s^2) - \bigg(\frac{k}{\lambda}\bigg)
(\lambda v_s)  A_k~,\nonumber\\
M_{45}&=& 2  \frac {k} {\lambda} \big[    (\lambda^2 v_s^2)  \sin 2 \beta \sin
\varphi_{ \lambda t}\big]~,\nonumber\\
M_{55}&=& \bigg( \frac{v^2}{ v_s^2 }\bigg)  \sin \beta \cos \beta \bigg[
m_A^2 \sin \beta \cos \beta +3  \bigg( \frac{k}{\lambda}\bigg)  
(\lambda^2 v_s^2)  \cos \varphi_{\lambda t} \bigg] \nonumber\\
&+& 3 \bigg( \frac{k}{\lambda}\bigg)  (\lambda v_s) A_k~, 
\end{eqnarray}
where 
\begin{eqnarray}
m_A^2&=&\frac {(\lambda v_s) [A_\lambda + k v_s \cos \varphi_{\lambda t}]}{\sin
\beta\cos\beta}~.
 \end{eqnarray}

The radiative correction terms due to the stop and sbottom corrections at the  
one-loop ($\Delta M_ {ij}$) can be found in the work of Ref. [32].
However, as mentioned before,  we differ from  Ref. [32]
in the sense that, in our analysis, the CP violating phase at the  tree-level  
is not equal to the one at the one-loop.

In our analysis, we will particularly concentrate  on the
lightest Higgs boson, whose mass can be obtained by the 
diagonalization of the  Higgs mass--squared matrix  by the similarity 
transformation: 
\begin{eqnarray}
{\cal{R}}M_{H}^{2}{\cal{R}}^{T}= {\rm diag}(m_{h_{1}}^{2},
m_{h_{2}}^{2}, m_{h_{3}}^{2},  m_{h_{4}}^{2}, m_{h_{5}}^{2})~,
\label{diagonal}
\end{eqnarray}
where ${\cal{R}}{\cal{R}}^{T}=1$. Here, we define $h_5$ to be the
lightest of all five  Higgs bosons .

The  mass eigenstates of 
the lightest 
Higgs boson ($h_5$) can then be decomposed 
in terms of the basis elements as:
\begin{eqnarray}
\label{Rij}
h_5=\sum_{i=1}^5 {{\cal{R}}_{5i} \Phi_{i}}~, 
\end{eqnarray}
where   $\Phi_{1}$, $\Phi_{2}$,  $\Phi_{3}$,   $\Phi_{4}$,   $\Phi_{5}$
correspond respectively,  $\phi_{1}$, $\phi_{2}$,
$\sin\beta \varphi_{1}+\cos\beta \varphi_{2}$, $\phi_s$ and $\varphi_s$
components of the Higgs boson under consideration.

From  Eq. (\ref{Rij}),  we define the dimensionless quantity $\rho_i$,
\begin{eqnarray}
\label{Rij2}
\rho_{i}=100\times |{\cal{R}}_{5i}|^{2} \,\,\, i=1, 2, 3, 4, 5~,
\end{eqnarray}
which is a measure of the percentage  CP component  
of a given mass--eigenstate  Higgs boson.
Therefore, in  (\ref{Rij2}) for instance, 
$\rho_{3}$ and  $ \rho_5$ 
are measures of the percentage CP odd components of 
$h_5$.

The main  contributions to the one-loop eEDM come from the  neutralino, and
chargino exchanges which can be expressed as:
\begin{eqnarray} 
\bigg(\frac{d_e}{e}\bigg)^{1-loop}&=&\bigg(\frac{d_e}{e}\bigg)^{\tilde
e-\chi^{0}_{i}} \ + \ \bigg(\frac{d_e}{e}\bigg)^{\tilde {\nu_{e}}-
\chi_{i}^{+}}~.
\label{oneloop}
\end{eqnarray}
By taking into account of the neutralino-selectron interaction, the neutralino contribution to the eEDM 
can be written as: 
\begin{eqnarray}
\label{neutralinochargino1}
\bigg(\frac{d_e}{e}\bigg)^{\tilde e-\chi^{0}_{i}}&=& \frac{\alpha } {4 \pi
 s^2_{W}}  \Bigg\{ \sum_{k=1}^{2} \sum_{i=1}^5 {{\it{Im}} [{\eta}_ {e_{ik}}] \ M_{{\chi}_{i}^{0}, {\tilde e}_{k}}
\ {\cal B} \big( M_{{\chi}_{i}^{0}, {\tilde e}_{k}}^2 \big)}\Bigg\}~, 
\end{eqnarray}
which is very similar to that of MSSM~\cite{Edm4}, except  
for the fact that the neutralino sector  now extends to  $5 \times 5$
mass matrix in the NMSSM, due to the presence of the additional gauge
singlet. By convention, in Eq. (\ref{neutralinochargino1}) and in the 
formalism below, we use the short-hand 
notation, for the generic indices 
$\alpha$ and $\rho$,  $M_{\alpha, \rho} \equiv \frac{ m_\alpha}{m_\rho}$,
and  $s_{W}(c_{W})=\sin{\theta_{W}}(\cos{\theta_{W}})$, \,
 $c_{\beta}=\cos{\beta}$, \,
$t_{W}=\tan{\theta_{W}}$.

In  Eq.( \ref{neutralinochargino1}) ${\eta}_{e_{ik}}$ is the neutralino 
vertex which is given by the following expression:
\begin{eqnarray}
\label{vertex}
{\eta}_{e_{ik}}&=&- \bigg[
\bigg (t_W {\cal{N}}_{1i}+ {\cal{N}}_{2i}  \bigg) \tilde{
{\cal{S}}}_{e1k }^{*}
+ \frac{M_{e,w} }{ c_{\beta} } {\cal{N}}_{3i} \tilde{ {\cal{S}}}_{e2k }^{*}
\bigg]\nonumber\\
&\times&
\bigg[t_W {\cal{N}}_{1i}  \tilde{ {\cal{S}}}_{e2k}
+\frac{M_{e,w}}{2  c_{\beta}} {\cal{N}}_{3i}  \tilde{ {\cal{S}}}_{e1k}
\bigg]~,
\end{eqnarray}
where ${\cal {N}}$ is the unitary matrix diagonalizing the neutralino matrix:
${\cal{N}}^{T} \,  M_{{\chi}^{0}} \,  {\cal{N}} =\mbox{diag}\left(m_{\chi^{0}_1}, \cdots,
m_{\chi^{0}_5}\right)$. The eigenstates $\left( \tilde{e}_1, \tilde{e}_2 \right)$
in Eq. (\ref{vertex}) can be obtained via the unitary rotation:
$\tilde{ {\cal{S}}}_{e }^{\dagger}\, \widetilde{M}^2_e\,
\tilde{{\cal{S}}}_{e} =
\mbox{diag}\left(m^{2}_{\tilde{e}_1},
m^{2}_{\tilde{e}_2}\right)$.

On the other hand, taking into account of the sneutrino-chargino interaction,
the  chargino contribution to eEDM reads as~\cite{Edm4}: 
\begin{eqnarray}
\label{neutralinochargino2}
 \bigg(\frac{d_e}{e}\bigg)^{\tilde {\nu_{e}}-
\chi_{i}^{+}}&=& \frac{\alpha } {4 \pi
 s^2_{W}}\Bigg\{
\frac  {M_{e,w}}{\sqrt{2} \ c_\beta  \ m_{{\tilde \nu}_{e}}^2}
\sum_{i=1}^2 { m_{  {\chi}_{i}^{+}}  { \it{Im}} [{\cal U}_{i2}^{*} {\cal V}_{i1}^{*}]
{\cal A} \big( M_{{\chi}_{i}^{+}, {\tilde \nu}_{e}}^2 \big)}\Bigg\}~,
\end{eqnarray}
where  ${\cal{U}}$ and  ${\cal{V}}$ are the unitary matrices diagonalizing the 
chargino mass matrix:
${\cal{U}}^{*} M_{C} {{\cal{V}}^{-1}} = \mbox{diag} (m_{ {\chi}_{1}^{+} }, \, m_{ {\chi}_{2}^{+}})$.
In Eqns. (\ref{neutralinochargino1})-(\ref{neutralinochargino2}),   
${\cal B}$ and ${\cal A}$ are the the loop functions~\cite{Edm4}.

We would like to note that  we 
take into account one loop contributions to  eEDM.
It was pointed out in Ref. [40] that in certain regions of MSSM parameter 
space certain two-loop contributions can not be non-negligible.
However, these two-loop contributions   become sizeable only at 
high $\tan\beta$ ($\tan\beta \simgt 30$). 
In this work we have restricted our attention 
to moderate $\tan\beta$ values ($\tan\beta \simlt 10$) which is a 
well-motivated parameter regime for the model under consideration~\cite{ellis1}. 
Hence the two-loop eEDM  will not provide significant 
contribution in our analysis.

Eq. (\ref{oneloop}), possesses the  
sources of CP violation through 

$(i)$ the gaugino masses
\begin{eqnarray}
(m_2, m_1) \rightarrow  (m_2 e^{i\varphi_2}, m_1 e^{i \varphi_1})~,  
\end{eqnarray}
which involves gaugino ($\varphi_1$) and the SU(2) gaugino ($\varphi_2$) 
phases in neutralino and chargino mass matrices.

$(ii)$ the complex selectron trilinear coupling
\begin{eqnarray}
(A_e) \rightarrow  (A_e  e^{i \varphi_e})~,
\end{eqnarray} 

Therefore, 
the phases of the stop and sbottom trilinear couplings
($\varphi_{A_{t}}=\varphi_{A_{b}}$), the phases of the 
gaugino masses ($\varphi_{1}$, and $\varphi_{2}$),
the  phases of  the  the Higgs potential, at tree and at the one-loop level
($\varphi_{k \lambda}$, and  $\varphi_{\lambda t}$, respectively)  
form the CP violating sources  in  the full parameter space.
In the following, we will perform a numerical study
to determine the eEDM, the effects of the physical phases of the
model on the mass  and 
on the  CP-odd components  ($\rho_{3}$ and $\rho_{5}$)  of $h_5$,
in the parameter space allowed by the eEDM constraint.

\section{Numerical Analysis}

In this section we will consider various parameter planes to adress the
issue of whether or not the various parameters
would lead to a large amount of CP violation opportunities.
As seen in the previous section, there are two physical phases which contribute to the
Higgs boson mass matrix. Besides these  physical  phases,
the free parameters appearing at the tree and one-loop levels are
$\tan\beta$, $A_k$, $A_\lambda$, $k$, $A_k$,  $v_s$.
$A_t$ and $A_b$.

In our  analysis, we particularly concentrate on 
$h_5$, and analyze  various parameter planes,
in the parameter space allowed by the  eEDM constraint.
In doing this, we use the present experimental upper bound of the 
eEDM~\cite{Commins,Abdullah}:
\begin{eqnarray}
d_e < 4.3  \times 10^{-27} \, \mbox{e.cm}~,
\end{eqnarray}

A convinient way to observe the effects of the eEDM
constraint, is via the dimensionless  quantity:
\begin{eqnarray}
\mbox{eEDM}&=& \frac{{\big[d_e/e\big]}^{th}}{\,\, {\big[d_e/e \big]}^{exp}}~,
\end{eqnarray}
which measures the fractional enhancement or suppression
of the eEDM with respect to its  experimental value.

In the numerical analysis, for the purpose of definiteness, we set:
$(i)$ the trilinear couplings  $A_t=A_b$
$(ii)$ $A_{\lambda}=v_s$, and $A_k=100~\mbox{GeV}$.
We focus on the values of  $k$,
and $\lambda$ that are favored by the renormalization group equations in the NMSSM~\cite{ellis1}.
In doing this, we fix the  $k$ parameter to be near its fixed point value
(i.e. we choose   $k=0.63$), and   we
focus on the allowed values of  $\lambda$,
in the $0.12 \simlt \lambda \simlt 0.82$ interval.

We  first  search  whether or not the  eEDM is consistent with the
present experimental bounds in the above mentioned intervals
provided that  the gaugino masses are of ${\cal O}(\mbox{TeV})$.
Namely,
we take:
$(i)$ the gaugino masses as $M_2=2000~\mbox{GeV}$,  $M_1=1000~\mbox{GeV}$,
$(ii)$ the slepton masses  from the neutralino and chargino sector as
$M_{\tilde L}=1500~\mbox{GeV}$,
and $M_{\tilde R}=1000~\mbox{GeV}$.

As mentioned in the Introduction, we take into account of not very large
values of $\tan\beta$ (i.e. $\tan\beta\simlt 10$).
For the purpose of showing the 
mass spectra, we shall take   $\tan\beta=2$. To give feeling of the 
sensitivity of the mass on $\tan\beta$, we shall also consider 
$\tan\beta=10$ regime.
In the numerical analysis, we concentrate on  two specific cases: 
in the first part, we carry out the analysis by letting $\lambda$ of a real parameter,
then in the second part we take into account of the case for which  $\lambda$
is complex.

\subsection{ The Case of Real $\lambda$}

In the first part of our analysis, we take $\lambda$ as a real parameter
($\varphi_{\lambda}=0$),  and let all the other phases in the theory of being complex. 
Namely, $\varphi_{A_t}$= $\varphi_{A_b}$=$\varphi_{A_e}$=$\varphi_{1}$=
$\varphi_{2}$=$\varphi$. 
In Fig. 1, we show the dependence of $|eEDM|$ on $\varphi$ and $\lambda$ 
for selected values of the vacuum expectation value of the singlet
($v_s$) when $\tan\beta=2$~(left panel) and  $\tan\beta=10$~(right  panel).
In the figure, we obtain 3-dimensional surfaces for each value of $v_s$,
which we choose within the $v \simlt v_s \simlt 4 v$
interval. For instance,   the top surface corresponds to  $v_s= 4 v$, 
whereas the bottom represents  $v_s= v$.
Here, $\varphi$ changes from 0 to $ 2 \pi$, and $\lambda$
from 0.12 to 0.82. 
As both panels of Fig. 1 suggest  the upper bound on  $|eEDM|$ gradually increases, 
with the increasing values of  $\lambda$, and  of  $v_s$.
One notes that the present bound of $|eEDM|$ is 
%satisfied in 
consistent with the present experimental bound 
in the full  $\varphi$ range, for all 
values  $v \simlt v_s \simlt 4v $, and  $ 0.12 \simlt \lambda \simlt 0.82$,
at both $\tan\beta=2$~(left panel)  and  $\tan\beta=10$~(right  panel) 
regimes, provided that  the gaugino masses are of ${\cal O}(\mbox{TeV})$. 
\begin{figure}[htb]
\vspace*{-1.6truein}
\centerline{\psfig{file=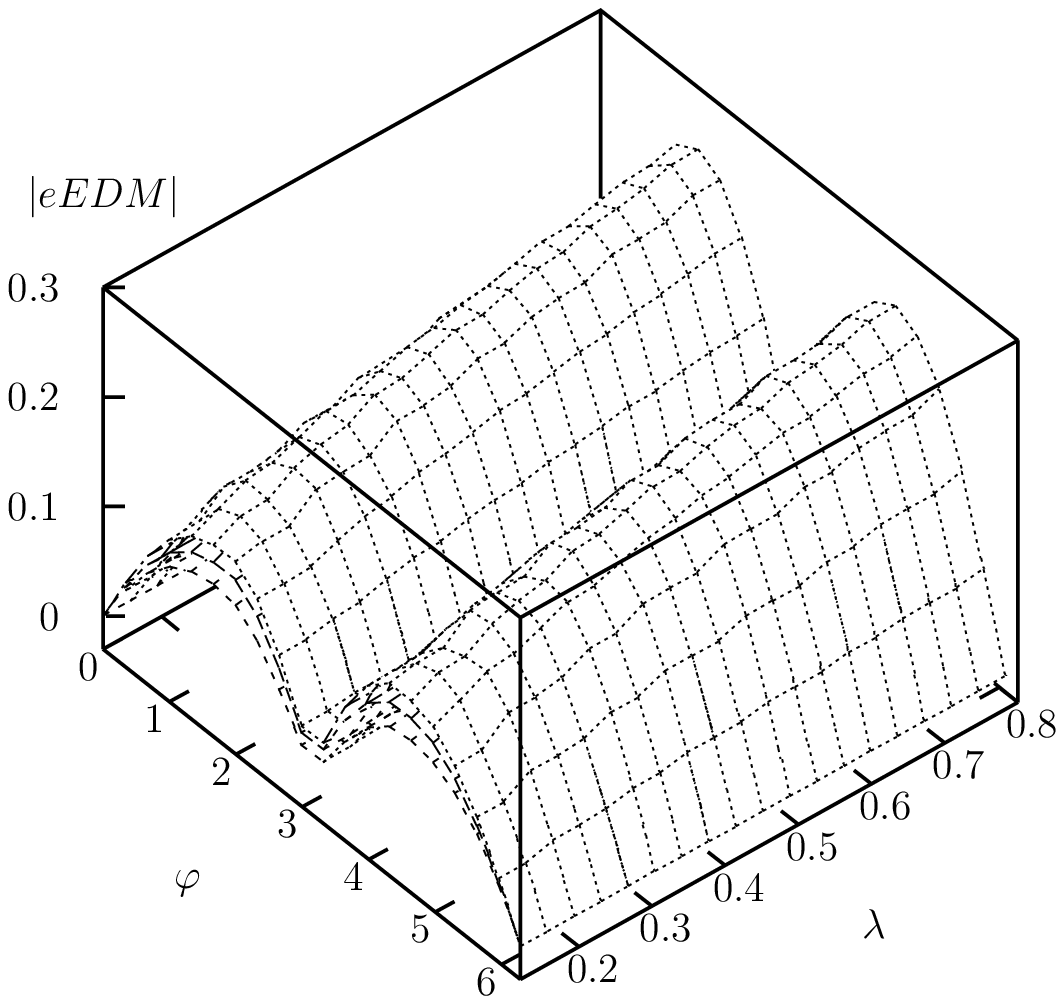,height=5.5in,width=5.0in }
\hspace*{-2.5truein}
\psfig{file=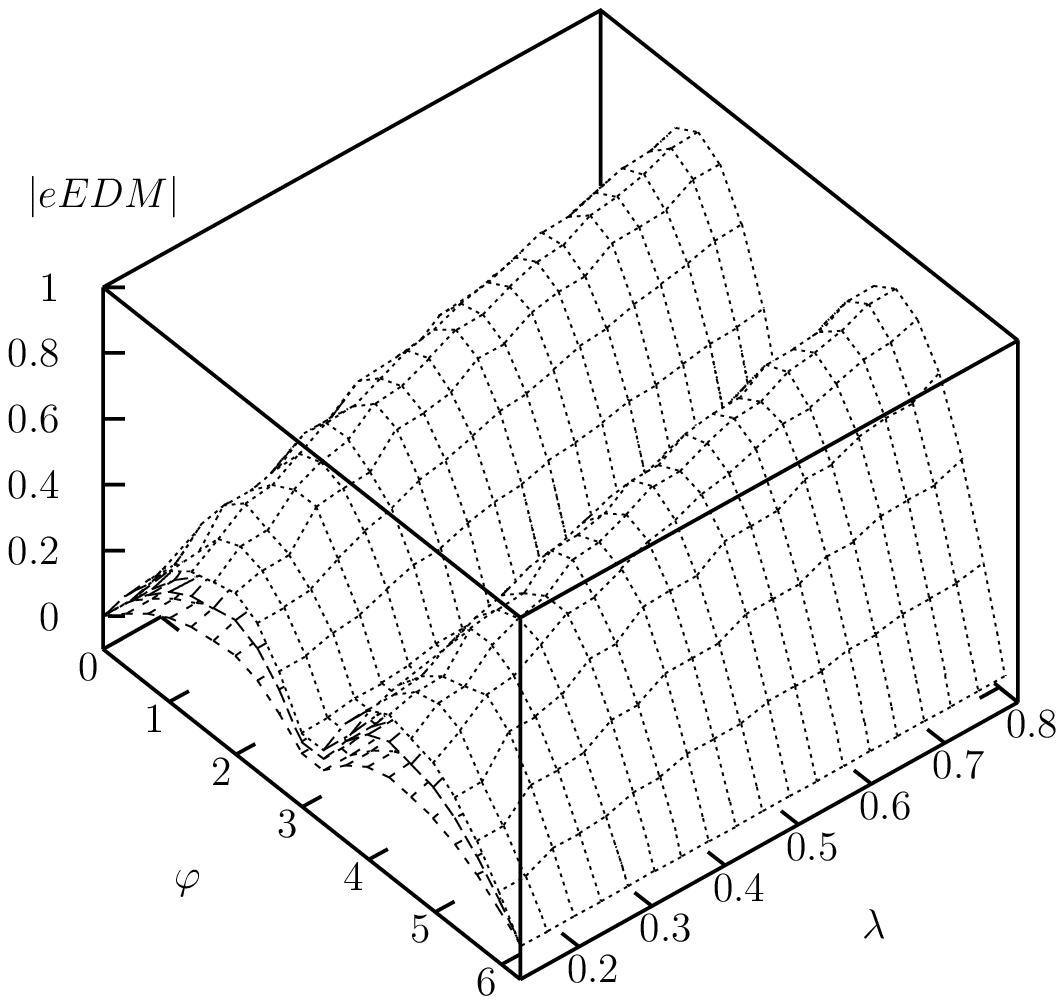,height=5.5in,width=5.0in }}
\vspace*{-1.7truein}
\fcaption{ The dependence of $|eEDM|$ on $\varphi$ and $\lambda$ 
for selected values of the vacuum expectation value of the singlet
when $\tan\beta=2$~(left panel) and  $\tan\beta=10$~(right  panel).}
\label{fig1}
\end{figure}

The analysis of  Fig. 1 gives a general idea of the
parameter domain of the   $|eEDM|$ and $\varphi$-$\lambda$ plane,
when $\varphi_{\lambda}=0$, and $\varphi$ changes in its full 
range. 
With this input in mind, we choose two values of 
$ \lambda$ ($\lambda=0.12$, and $\lambda=0.45$),
and  of $v_s$ ($v_s= v$ and $v_s= 3 v $), in the parameter space allowed by the
eEDM constraint, corresponding to the low and high values of 
$\lambda$ and $v_s$, respectively, to analyze the dependence of the mass
($m_{h_{5}}$) at the tree and
one-loop levels, and of the CP odd components of the lightest Higgs boson
($h_5$) on $\varphi$ at both $\tan\beta$ regimes. 

We would like to note that,
in the remaining part of the analysis 
the variation of   
$m_{h_{5}}$
as well as $\rho_3$ and $\rho_5$,  with $\varphi$
is displayed 
for several values of 
$A_t$
taking into account of the  strong dependence of the 
radiative corrections on the stop splitting, as will be seen in Eq. (22).

Therefore, in each  of the following plots, we first consider the case 
for which $A_t$ and $v_s$ are  of comparable size,
(i.e, $A_t=v_s$) when  $v_s=175~\mbox{GeV}$,  and $v_s=525~\mbox{GeV}$.
Next, to determine how the increase  in $A_t$  affects the 
radiative corrections, 
we concentrate on  two specific values of $A_t$  
corresponding to   $A_t=1050\mbox{GeV}$,
and   $A_t=1400\mbox{GeV}$  for  
both 
$v_s=175~\mbox{GeV}$ and $v_s=525~\mbox{GeV}$ cases.

In Fig. 2, we  show  the dependence  of   $m_{h_{5}}$  
(at  tree and at one-loop levels) on $\varphi$  at   
$\tan\beta=2$~(left panel) and   
$\tan\beta=10$~(right panel), when $v_s= v =175~\mbox{GeV}$,
and   $\lambda=0.12$, for selected values of  $A_t$. 
In both panels of Fig. 2, the three
upper curves with respect to the mid-point,  from bottom to top, 
represent  $A_t=v_s=175~\mbox{GeV}$,  
$A_t=1050~\mbox{GeV}$,
and   $A_t=1400~\mbox{GeV}$
values of $A_t$ respectively, whereas the  lowest curve is for  
the tree-level.
\begin{figure}[htb]
\vspace*{-2.1truein}
\centerline{\psfig{file=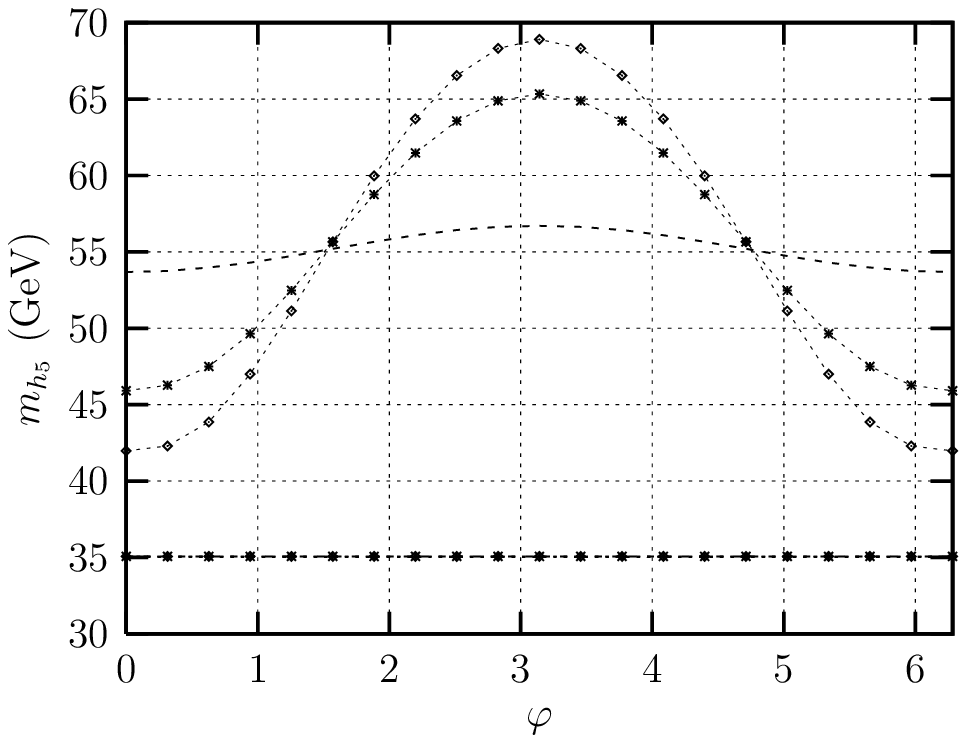,height=6in,width=4.8in }
\hspace*{-2.5truein}
\psfig{file=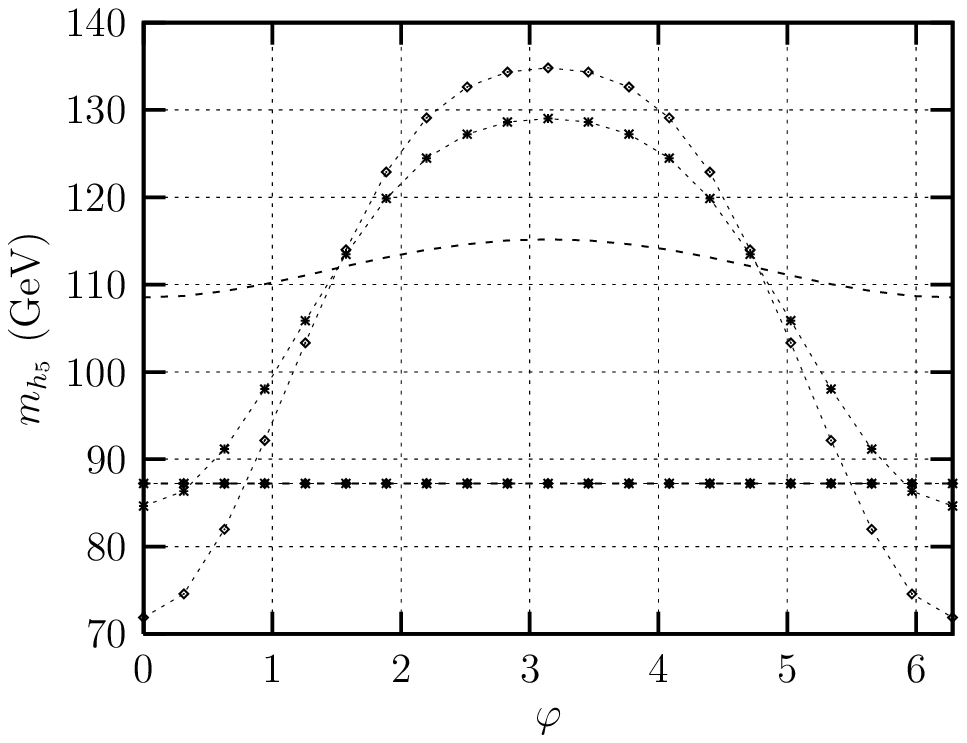,height=6in,width=4.8in }}
\vspace*{-2.2truein}
\fcaption{ The dependence of  
$m_{h_{5}}$
on $\varphi$, for selected values of $A_t$,
at $\tan\beta=2$~(left panel), and $\tan\beta=10$~(right panel)
when  $v_s=175~\mbox{GeV}$  and  $\lambda=0.12$. 
Here, the three upper curves with respect to the mid-point,  
from bottom to top, 
are for $A_t=v_s=175~\mbox{GeV}$, 
$A_t=1050~\mbox{GeV}$, and   
$A_t=1400~\mbox{GeV}$, 
respectively, whereas the  lowest curve 
is for the tree-level.}
\label{fig2}
\end{figure}

As Fig. 2 suggests  when   $\lambda=0.12$, and $v_s=175~\mbox{GeV}$,
$m_{h_{5}}$ grows from 
(42,\,46,\,54) to  (69,\,65,\,56)~$\mbox{GeV}$ at  $\tan\beta=2$,
and from (71,\,85,\,109)  to  (135,\,130,\,115)\,\,\,
$\mbox{GeV}$ at  
$\tan\beta=10$,
for the values of $A_t=1400~\mbox{GeV}$~(the top curve with respect to the  mid-point),  
$A_t=1050~\mbox{GeV}$~(the second curve below the top curve), and
$A_t=v_s$ (the third curve below the top curve), respectively, as $\varphi$ 
ranges from 0 to $\pi$.   On the other hand, 
remaining around  $35$ $\mbox{GeV}$ for  $\tan\beta=2$,      
$m_{h_{5}}$ grows up  until  $87$ $\mbox{GeV}$,  at the tree level, for   $\tan\beta=10$. 
It can also be observed from Fig. 2 that  
$m_{h_{5}}$ is quite sensitive to the variations in $\varphi$ for 
$\lambda=0.12$, and  $v_s=175~\mbox{GeV}$
for both   $\tan\beta=2$~(left panel) and   $\tan\beta=10$~(right panel)
regimes. One notes that the splittings between the tree and one-loop spectra are
quite small. However,   as  $v_s$ gets higher values, 
the strength of the radiative corrections 
are affected, which causes larger splittings between  the tree
and one-loop mass spectra. 
\begin{figure}[htb]
\vspace*{-2.1truein}
\centerline{\psfig{file=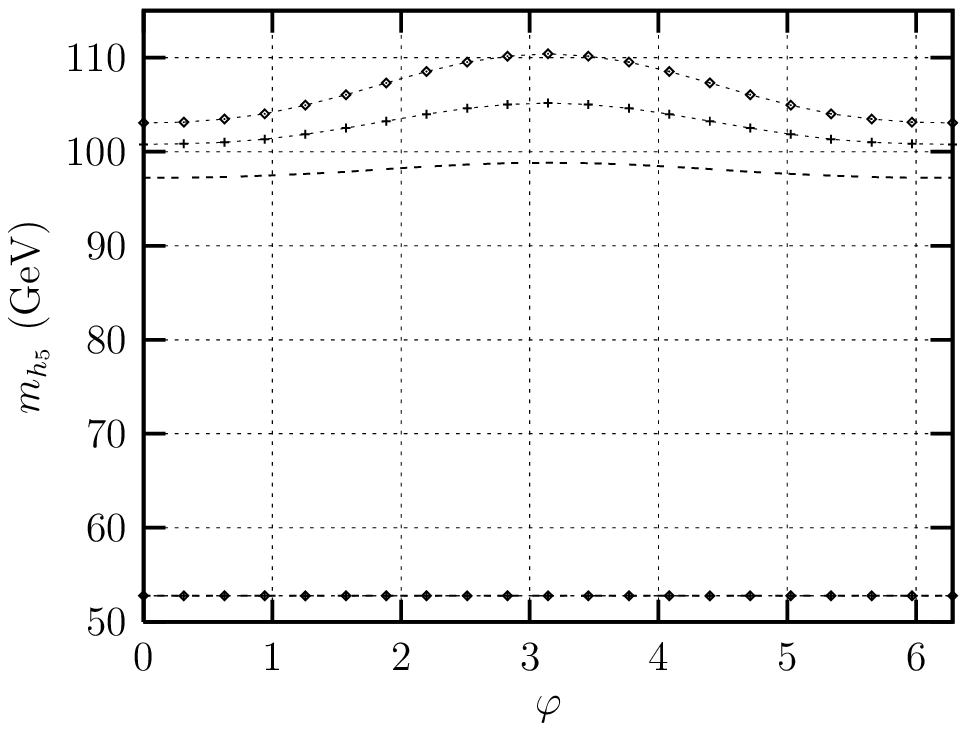,height=6.0in,width=4.8in }
\hspace*{-2.5truein}
\psfig{file=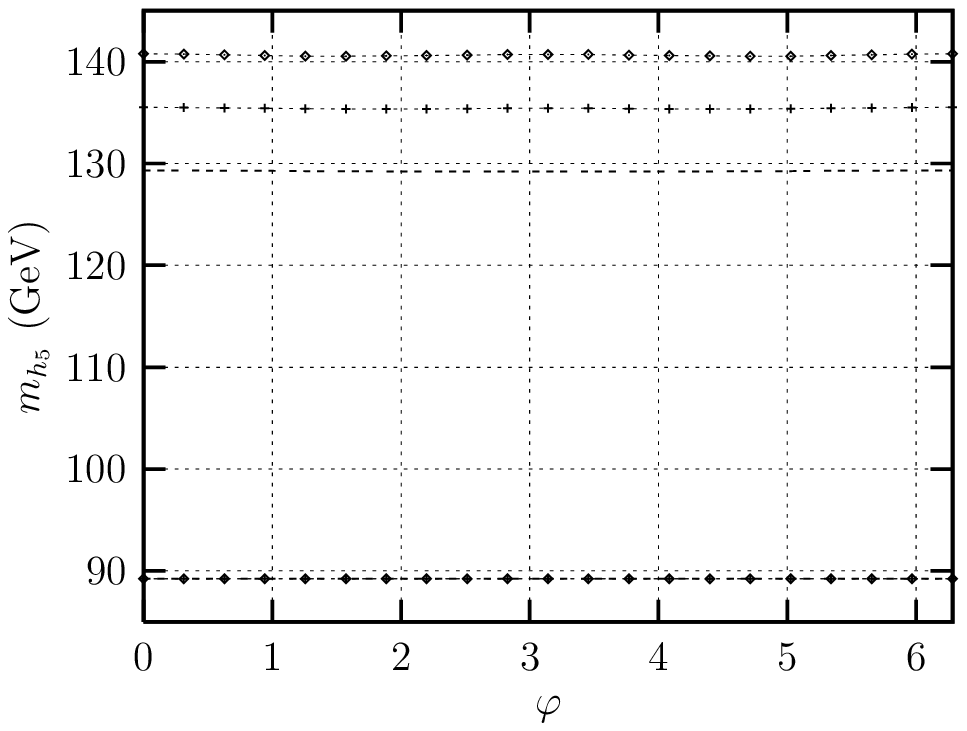,height=6.0in,width=4.8in }}
\vspace*{-2.2truein}
\fcaption{ The dependence of  
$m_{h_{5}}$
on $\varphi$, for 
selected values of $A_t$, at 
$\tan\beta=2$~(left panel), and $\tan\beta=10$~(right panel), 
when  $v_s$=525~$\mbox{GeV}$  and  $\lambda$= 0.12.  
Here, the  bottom, the middle and the top curves,
are for $A_t=v_s=525~\mbox{GeV}$,
$A_t=1050~\mbox{GeV}$, and   $A_t=1400~\mbox{GeV}$ values, 
whereas the  lowest curve is for the tree level.} 
\label{fig3}
\end{figure}

For instance, in Fig. 3, we show  the dependence  of  
$m_{h_{5}}$ on  $\varphi$, for   
 $v_s= 3 v =525~\mbox{GeV}$,
and $\lambda=0.12$. 
In Fig. 3, we select three specific values of $A_t$: namely 
$A_t=1400~\mbox{GeV}$~(the top curve),
$A_t=1050~\mbox{GeV}$~(the second curve below the top curve),
$A_t=v_s~\mbox{GeV}$~(the third curve below the top  curve)
Here, the  lowest curve represents  $m_{h_{5}}$ at the
tree-level. It can be seen from the left panel of Fig. 3 that 
as $v_s$ increases the gap 
between the tree and the one-loop levels enlarges. For instance, being around 
at most $\sim 20$ $\mbox{GeV}$ for $v_s=v$ and $A_t=v_s$~(left panel of Fig. 2), it rises to $\sim 50$
$\mbox{GeV}$ for $v_s=3 v$  at $\tan\beta=2$~(left panel of Fig. 3). 
This behaviour occurs also in the   $\tan\beta=10$ regime for which case 
the gap between the tree and one-loop levels increases (right panel of Fig. 3).

A comparative analysis of left panels of Fig. 2 and  Fig. 3 
shows that
when $A_t$ and $v_s$ are of comparible size,
the
variation of $m_{h_{5}}$ 
with $\varphi$  is much more slower as compared to 
the cases for which $A_t >  v_s$ 
($A_t=1050~\mbox{GeV}$, and   $A_t=1400~\mbox{GeV}$).
This saturation effect can be understood 
by observing that the radiative corrections depend strongly on the 
stop splitting  $\Delta_{\tilde{t}}^{2}$, which depends explicitely on $\varphi_{\lambda t}$ 
such that
\begin{eqnarray}
\delta=\Delta_{\tilde{t}}^{2}(\pi)  /  \Delta_{\tilde{t}}^{2}(0)=
\frac{|A_t|-|\lambda| v_s \cot\beta} {|A_t|+|\lambda| v_s \cot\beta}
=\frac{1-\frac{|\lambda|\cot\beta }{ |A_t|/v_s}} 
{1+ \frac{|\lambda|\cot\beta} {|A_t|/v_s}}~.
\label{delta}
\end{eqnarray}
This quantity particularly implies that the strength of the radiative
corrections modify from one CP conserving point 
to the next. It can also be seen from (\ref{delta}) 
that when $|A_t|$ and  $v_s$, are of comparible size,
as $\tan\beta$ increases,
$|\lambda|  \cot\beta$
decreases
and  indeed, $\delta$
approaches  to unity in the high $\tan\beta$ limit. 
One notes that  the tree level mass of the model
is proportional to 
$\lambda^2 v_s^2$, whereas 
the CP violating entries 
of the radiative corrections 
of the Higgs mass-squared matrix 
grow with the term $|A_t| v_s \lambda$.
Naturally, with the  increase in $A_t$,  $|A_t| v_s \lambda$,
term also increases. That is  
the radiative corrections enhance through the 
$A_t$ term, which does not cause too big splittings 
among the radiative corrections, 
when   $v_s=v$ and $\lambda=0.12$.
However, as $v_s$
gets higher values, the seperation from the tree-level enlarges.

In Fig. 4, we show the dependence of  CP-odd components 
($\rho_3$ and $\rho_5$) of $h_5$  on $\varphi$, for $\lambda=0.12$, and    
$v_s=v=175~\mbox{GeV}$,    
when   $\tan\beta=2$~(left panel) and   $\tan\beta=10$~(right panel).
In both panels the upper and the lower curves represent the $\rho_3$, and 
$\rho_5$ components of $h_5$, respectively.
We select three values of  $A_t$:
$A_t=v_s=175~\mbox{GeV}$,  
$A_t=1050~\mbox{GeV}$,  $A_t=1400~\mbox{GeV}$,
from top to bottom for the upper curves ($\rho_3$),
and from bottom to top for  the lower curves ($\rho_5$). 
\begin{figure}[htb]
\vspace*{-2.1truein}
\centerline{\psfig{file=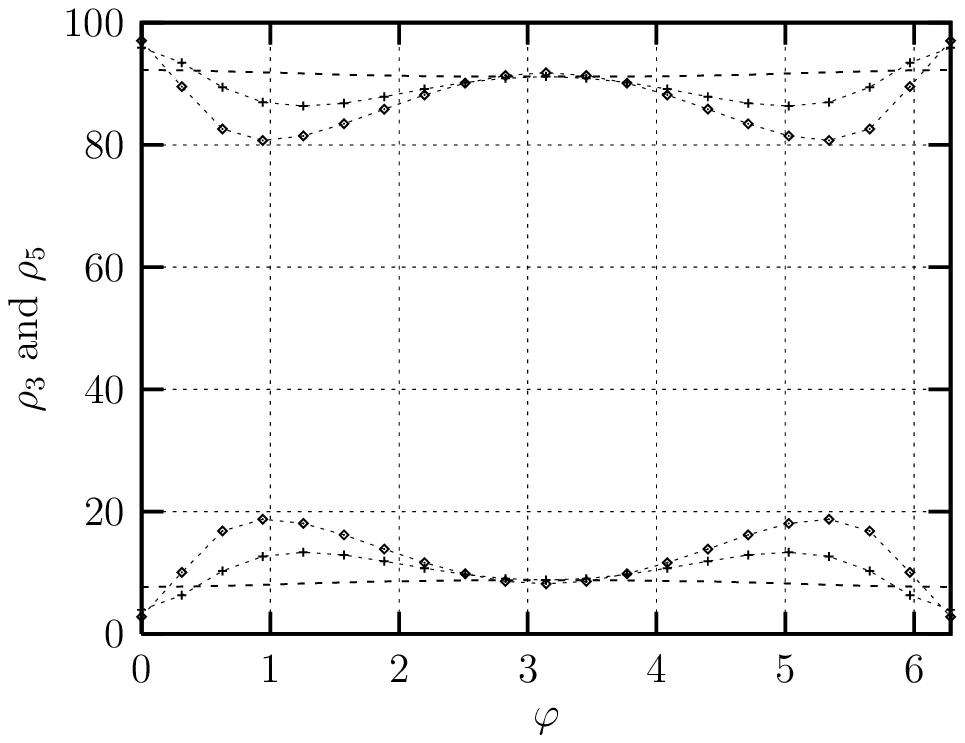,height=6.0in,width=4.8in }
\hspace*{-2.5truein}
\psfig{file=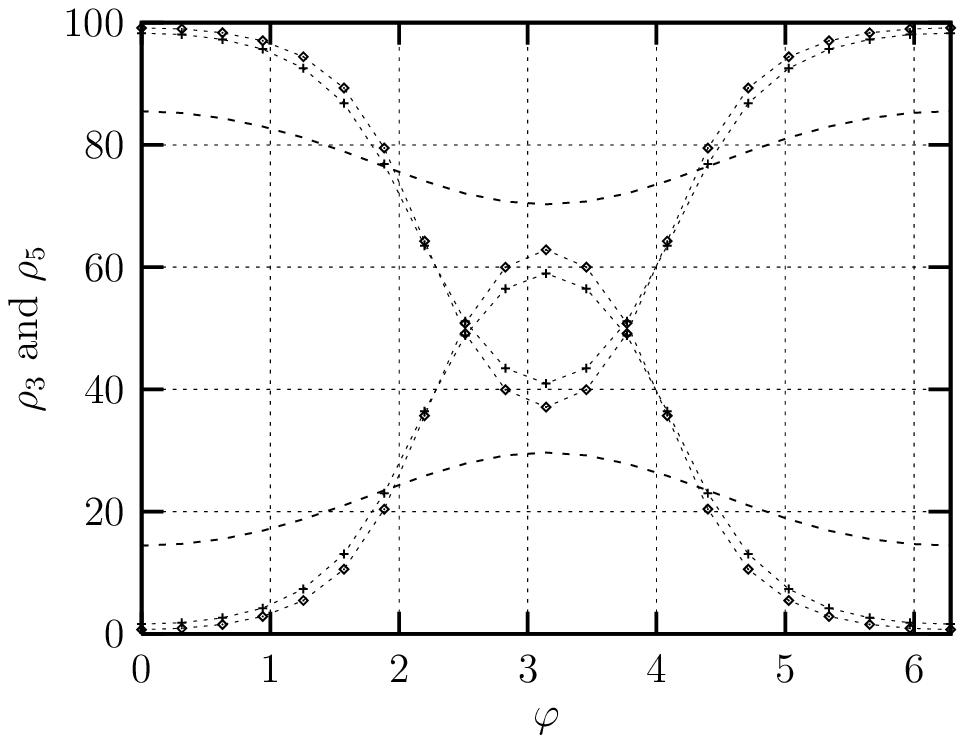,height=6.0in,width=4.8in }}
\vspace*{-2.2truein}
\fcaption{The dependence of
$\rho_3$~(the upper curves) and $\rho_5$~(the lower curves) on $\varphi$ for  
$v_s$=$175~\mbox{GeV}$, 
and $\lambda$= 0.12, for selected values of $A_t$,
when $\tan\beta=2$~(left panel), and $\tan\beta=10$~(right panel).
Here,   $A_t=v_s=175~\mbox{GeV}$, 
$A_t=1050~\mbox{GeV}$,  
$A_t=1400~\mbox{GeV}$, 
from top to bottom for $\rho_3$, and from bottom to top for  $\rho_5$.} 
\label{fig4}
\end{figure}

As the  left panel of Fig. 4 suggests,   
when   $A_t=1400~\mbox{GeV}$, and $A_t=1050~\mbox{GeV}$ 
$h_5$ has $\%98$  and  $\%96$ 
$\rho_3$ components, respectively, 
at $\tan\beta=2$ and $\varphi=0$.
When  $A_t=v_s$,   $\rho_3$ 
decreases to $\%92$.
The left panel of the Fig. 4 also indicates  that  
$\rho_5$   ranges in between  $\%2$ and  $\%20$, depending on 
the strength  of 
$A_t$.
On the other hand,  as we observe from the right panel,  when   $\tan\beta=10$, 
$\rho_3$  rises nearly to 
the $\%100$ line for  $A_t=1050~\mbox{GeV}$~(and  for $A_t=1400~\mbox{GeV}$, as well), 
and correspondingly  $\rho_5$ 
remains in the vicinity of the 
$\%0$  line  at $\varphi=0$. 
For higher values of $v_s$,
the $\rho_5$ component of $h_5$ increases,
and this increase in the  $\rho_5$ is compensated by $\rho_3$,
as expected (see Fig. 5). 
\begin{figure}[htb]
\vspace*{-2.1truein}
\centerline{\psfig{file=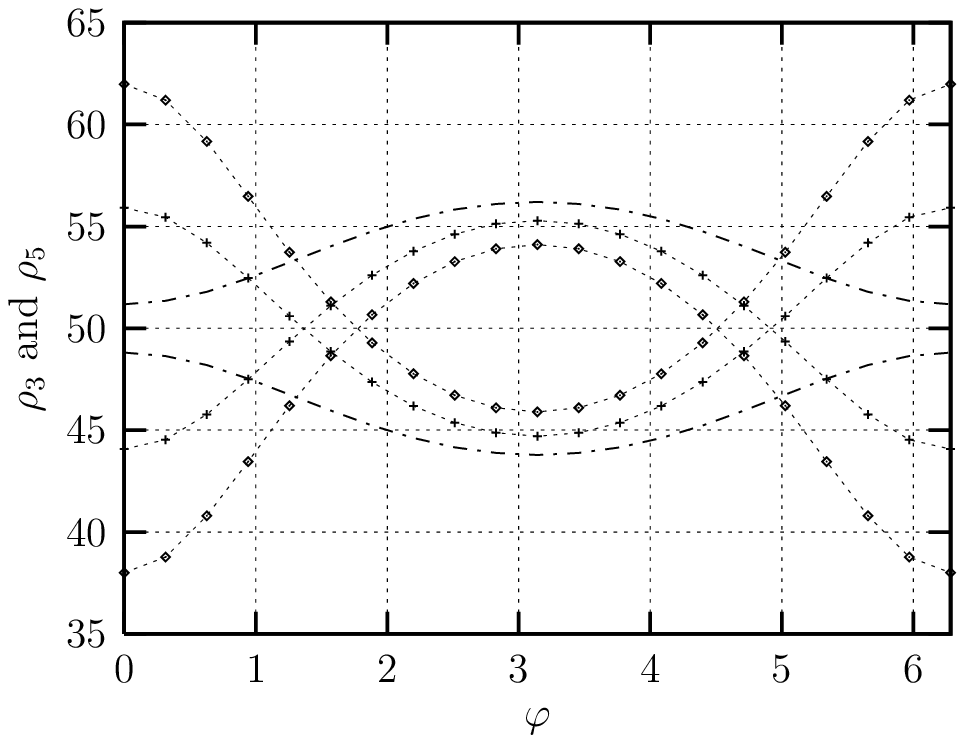,height=6.0in,width=4.8in }
\hspace*{-2.5truein}
\psfig{file=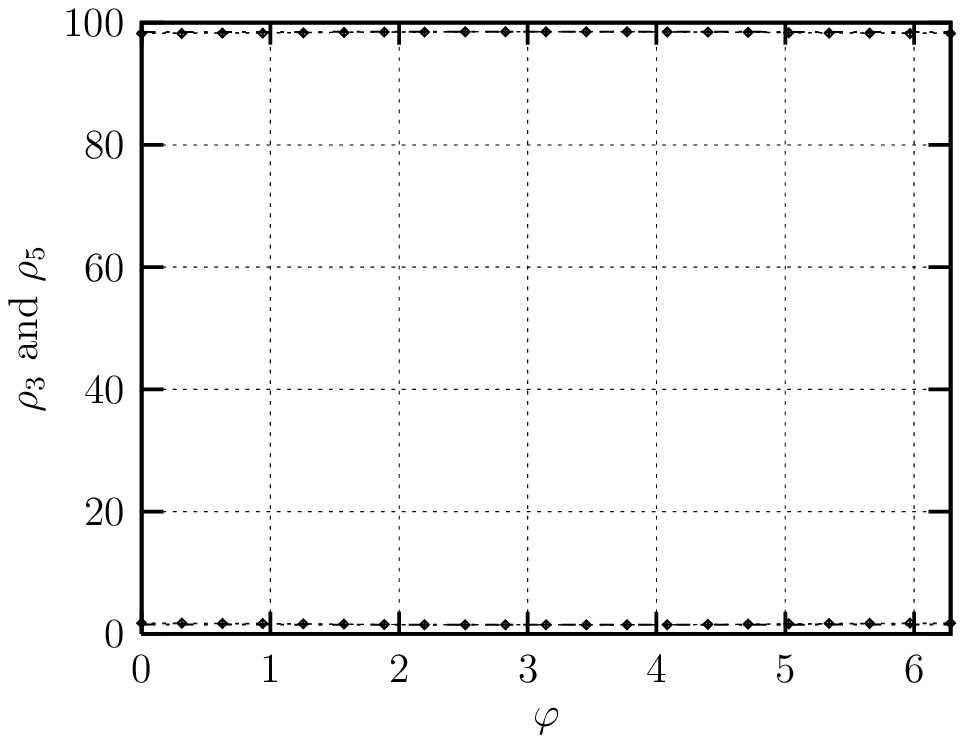,height=6.0in,width=4.8in }}
\vspace*{-2.3truein}
\fcaption{ The dependence of  $\rho_3$~(the lower curves) and 
$\rho_5$~(the upper curves) on $\varphi$ for  $v_s$= 525~$\mbox{GeV}$,  
and $\lambda$= 0.12, for selected values of $A_t$,
when $\tan\beta=2$~(left panel) and  $\tan\beta=10$~(right  panel).
Here, $A_t=v_s=525~\mbox{GeV}$, $A_t=1050~\mbox{GeV}$ and   
$A_t=1400~\mbox{GeV}$, 
from bottom to top for $\rho_3$, and  from top to bottom for  $\rho_5$, with
respect to the mid-point.}
\label{fig5}
\end{figure}

In Fig. 5, we show the dependence of  $\rho_3$ and  $\rho_5$ components of
$h_5$ on $\varphi$, for $\lambda=0.12$, and    
$v_s= 3 v=525~\mbox{GeV}$,   when   
$\tan\beta=2$~(left panel), and  $\tan\beta=10$~(right  panel). 
Here, the lower and the upper curves represent the $\rho_3$, and 
$\rho_5$ components of $h_5$, respectively.
Similar to the Fig.  4, in Fig. 5,  we select three values of  $A_t$:
$A_t=v_s$,
$A_t=1050~\mbox{GeV}$,  $A_t=1400~\mbox{GeV}$, 
from bottom to top for the lower curves ($\rho_3$), and from top to bottom
for the upper curves ($\rho_5$), with respect to the mid-point. 
It  can be noted from the left panel of Fig. 5 that  
as $v_s$  gets higher values the  $\rho_5$ component of  $h_5$
increases, with the decrease of its $\rho_3$
component, as compared to the previous case (see Fig. 4). For instance, 
when  $\tan\beta=2$~(left panel of Fig. 5), and   $A_t=1400~\mbox{GeV}$,
$\rho_3$ decreases to   
$\sim \%63$ whereas  $\rho_5$  increases to $ \sim \%37$
at $\varphi=0$. A more spectacular side of Fig. 5 arises for
$\tan\beta=10$~(the right  panel) where 
one observes a very slow variation of $\varphi$
as compared to the previous case (the right panel of Fig. 4).

A comparative analysis of Figs. 4 and 5 suggest that 
for small $\lambda$~($\lambda=0.12$)  and higher $v_s$ ($v_s=3 v$),   
the  $\rho_5$ component of  $h_5$  increases, 
and this increase is compensated by the decrease in  $\rho_3$.
Such kind of effect can be  dominantly seen in the high  $\tan\beta$
regime (see the right panel of Fig. 5).

One notes that in the MSSM there is  a single CP-odd component,  
and naturally the results of the CP-invariant 
theory are expected to be recovered at the CP-conserving points 
($\varphi=0, \pi, 2 \pi$)~\cite{Demir2001}.
However, there are two CP odd components in the NMSSM~($\rho_3$ and $\rho_5$),
and  these two odd components mix each other at the CP conserving points,
depending on the relative strengths of the other one-loop corrections.
Therefore, one necessarily does not recover the results of the CP-invariant 
theory
at the CP conserving points, due to the CP violating mixings of  
$\rho_3$ and $\rho_5$.

Until now, we have carried out our analysis  
for small values of   $\lambda$,  when $\varphi_{\lambda}=0$,
and  all the other phases in the theory are  assumed to be complex. 
In the following, we take into account of  higher values of   $\lambda$. 
For this reason, we  fix  $\lambda=0.45$ and carry out the similar 
analysis for the second part of the work.
\begin{figure}[htb]
\vspace*{-2.1truein}
\centerline{\psfig{file=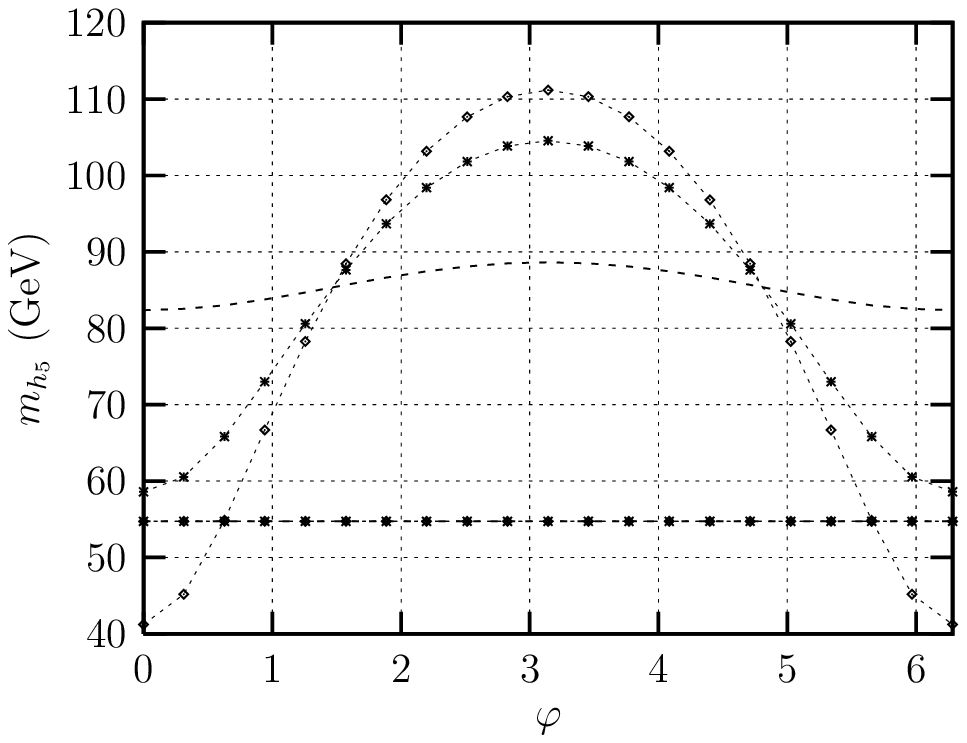,height=6.1in,width=4.9in }
\hspace*{-2.5truein}
\psfig{file=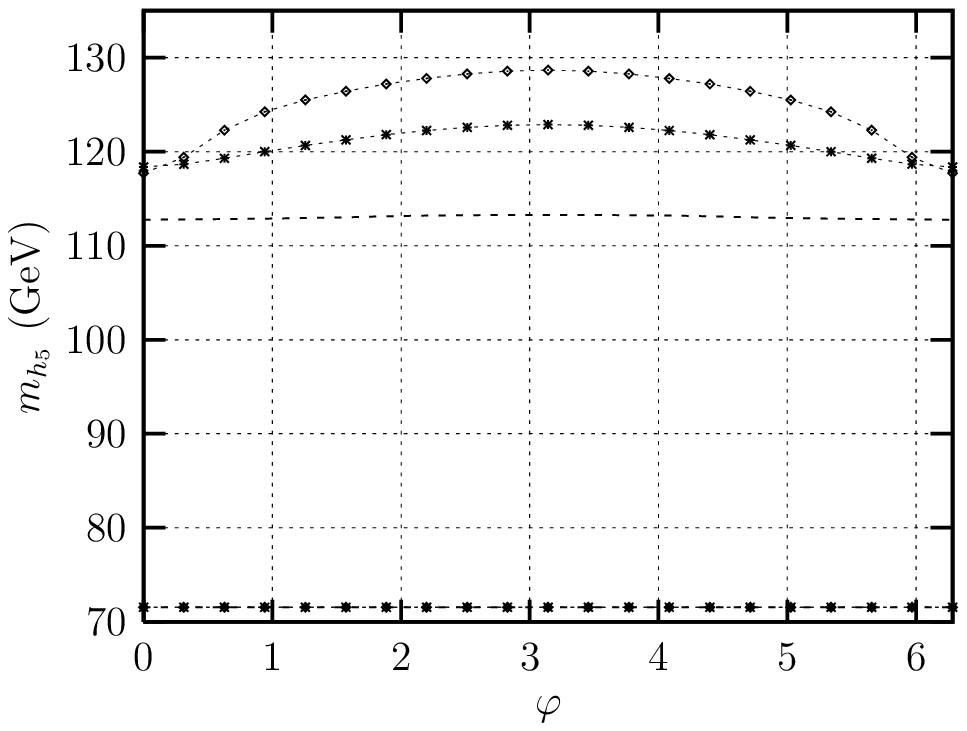,height=6.1in,width=4.9in }}
\vspace*{-2.3truein}
\fcaption{ The dependence of  
$m_{h_{5}}$
on $\varphi$, for selected values of $A_t$,
at $\tan\beta=2$~(left panel), and $\tan\beta=10$~(right panel)
when  $v_s$= v and  $\lambda=0.45$. Here, the three upper curves
with respect to the mid-point,  from bottom to top, 
are for $A_t=v_s=175~\mbox{GeV}$,
$A_t=1050~\mbox{GeV}$, and   
$A_t=1400~\mbox{GeV}$, 
respectively, whereas the lowest curve 
is for the tree level.}
\label{fig6}
\end{figure}

In Fig. 6, we show  the dependence  of   $m_{h_{5}}$  
(at  tree and at one-loop levels)
on $\varphi$
for    $v_s= v=175~\mbox{GeV}$,  and  $\lambda=0.45$, when   
$\tan\beta=2$~(left panel) and   $\tan\beta=10$~(right panel).
We again select three values of  $A_t$, namely 
$A_t=1400~\mbox{GeV}$~(the top curve with respect to the mid-point),
$A_t=1050~\mbox{GeV}$~(the second curve below the top curve),
$A_t=v_s=175~\mbox{GeV}$~(the third curve below the top  curve).
Here,  the  lowest curve is for the tree-level.
It can be observed  from Fig. 6 that  $m_{h_{5}}$ 
still remains sensitive to the variations in $\varphi$ for 
higher values of $\lambda$ ($\lambda=0.45)$ at  $\tan\beta=2$. 
As $\tan\beta$
increases, the dependence of 
top squark masses on 
$\lambda$, as well as  $v_s$ weakens, 
therefore these elements of the mass-squared matrix
become more sensitive to the choice of 
these parameters. Hence,  for  $\tan\beta=10$ (right panel of Fig. 6), 
one observes  a slow variation  with $\varphi$,
and quite large splitting between the tree and loop levels
as compared to $\lambda=0.12$ case (see Fig. 2, right panel).
Similar analysis can be performed for $v_s \simgt 2 v$,
and $\lambda=0.45$,
in which case one observes  a weak dependence of $\varphi$ on 
$m_{h_{5}}$ 
at both    $\tan\beta=2$ and  $\tan\beta=10$ regimes
(see Fig. 7). 
\begin{figure}[htb]
\vspace*{-2.1truein}
\centerline{\psfig{file=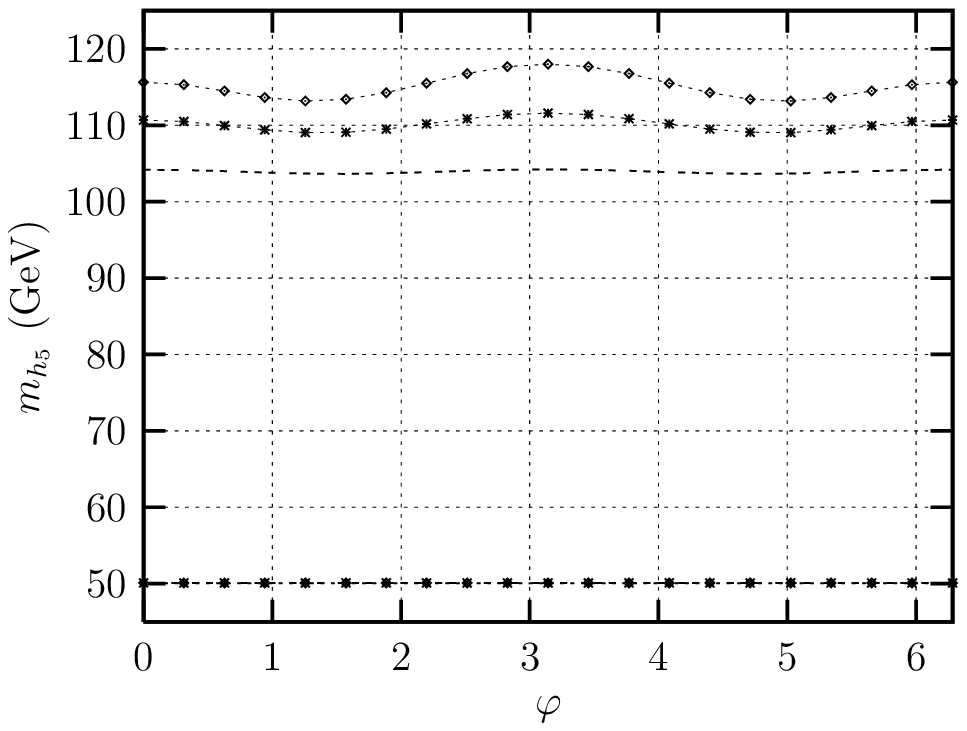,height=6.0in,width=4.9in }
\hspace*{-2.5truein}
\psfig{file=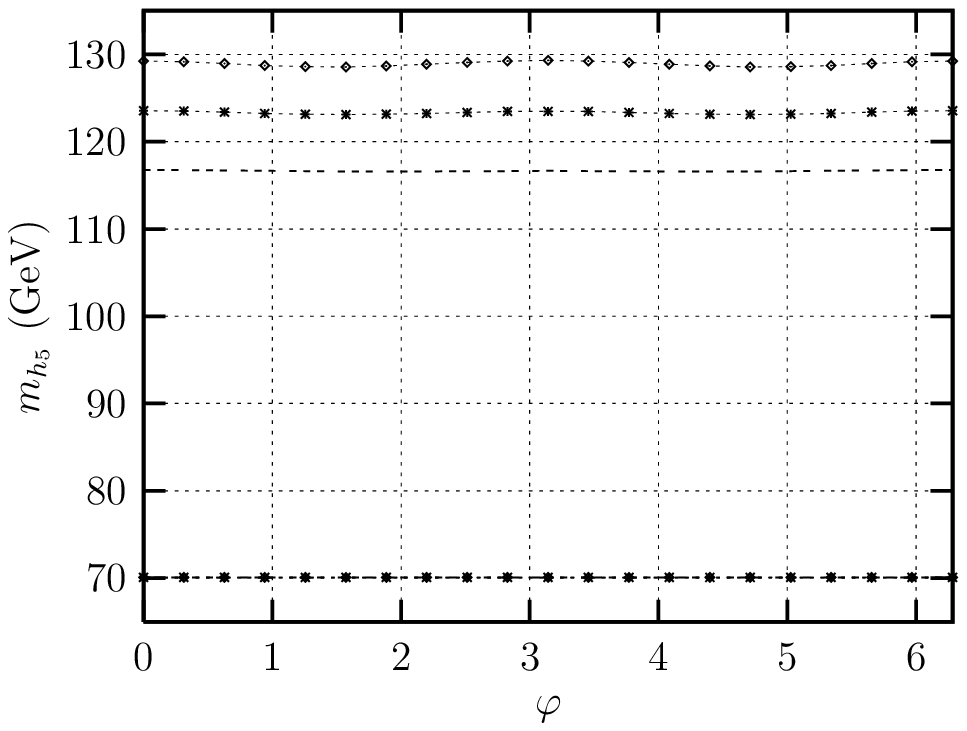,height=6.0in,width=4.9in }}
\vspace*{-2.3truein}
\fcaption{ The dependence of  
$m_{h_{5}}$
on $\varphi$ for  
$v_s$= 525~$\mbox{GeV}$, and $\lambda$= 0.45, for selected values of $A_t$,
when $\tan\beta=2$~(left panel), and $\tan\beta=10$~(right panel).
Here, the bottom, middle and the top curves 
are for $A_t=v_s=525~\mbox{GeV}$, $A_t=1050~\mbox{GeV}$, 
and   $A_t=1400~\mbox{GeV}$, 
respectively, whereas the  lowest curve 
is for the tree-level.}
\label{fig7}
\end{figure}

One notes from the left panel Fig. 7 that the increase in  
$v_s$ affects the radiative corrections,
which causes quite large splittings between the tree and the loop levels,
at $\tan\beta=2$.  On  the other hand,   
it is seen from the right panel that when  $\tan\beta=10$,
the variation of $\varphi$ is much more flat, 
as compared to right panel of Fig. 6.
\begin{figure}[htb]
\vspace*{-2.1truein}
\centerline{\psfig{file=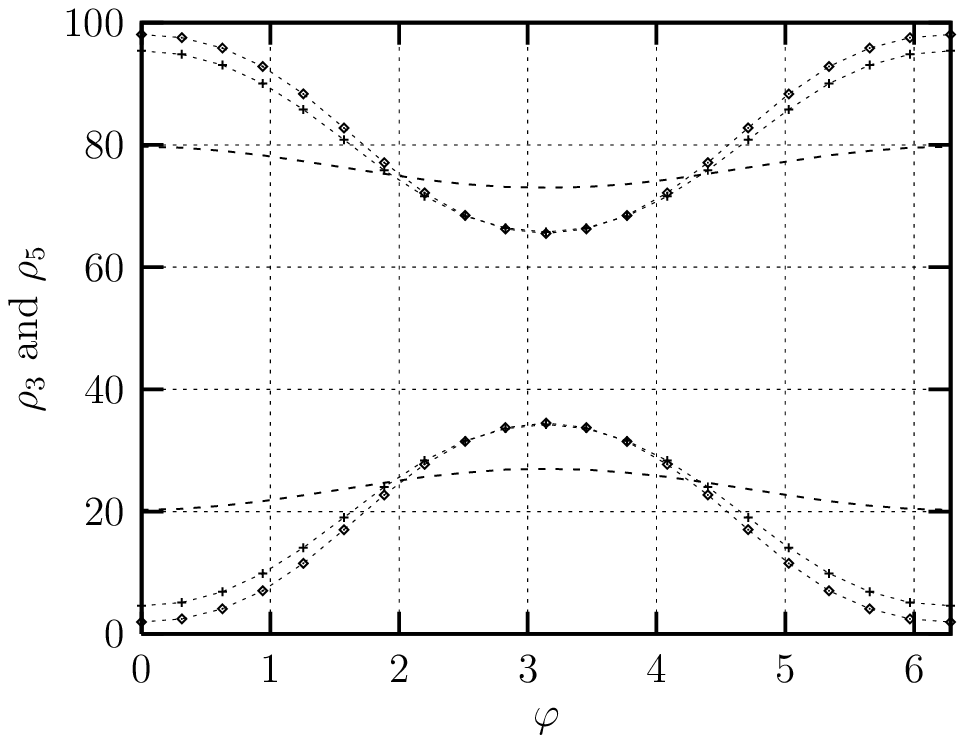,height=6.0in,width=4.9in}
\hspace*{-2.5truein}
\psfig{file=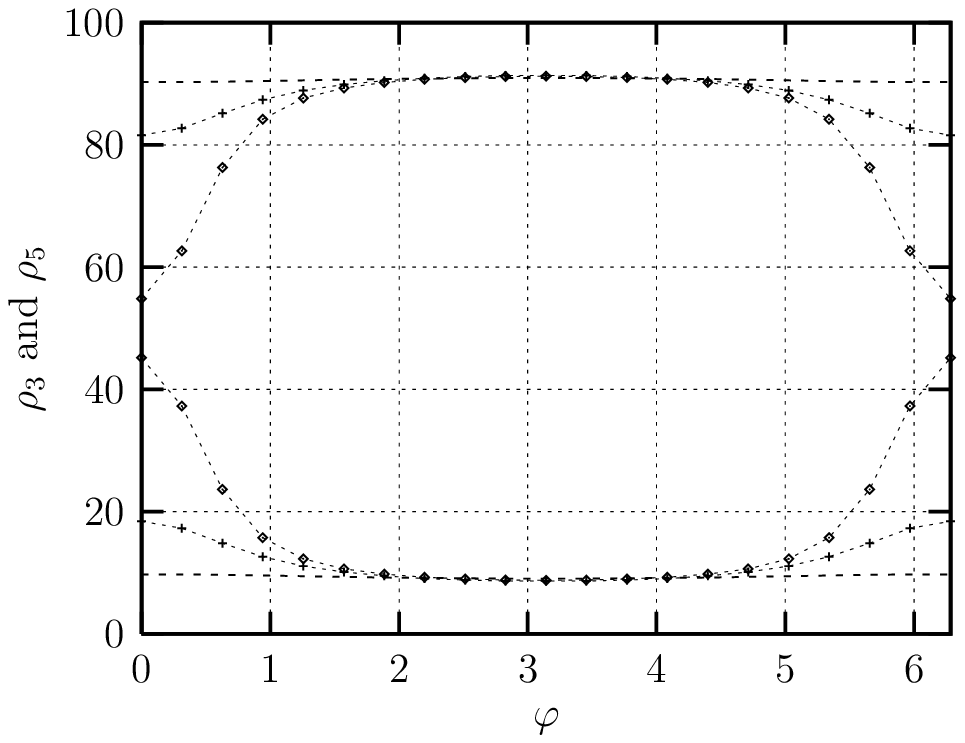,height=6.0in,width=4.8in }}
\vspace*{-2.3truein}
\fcaption{The dependence of 
$\rho_3$  and $\rho_5$ on $\varphi$ for  $v_s$=v, 
and $\lambda$= 0.45, for selected values of $A_t$,
when $\tan\beta=2$~(left panel), and $\tan\beta=10$~(right panel).
The upper (lower) curves are for $\rho_3$($\rho_5$) in the left panel,
whereas they are for  $\rho_5$($\rho_3$)  in the right  panel.
Here,   $A_t=v_s=175~\mbox{GeV}$, 
$A_t=1050~\mbox{GeV}$,  $A_t=1400~\mbox{GeV}$,
from top to bottom for $\rho_3$ ($\rho_5$), with respect to the mid-point.} 
\label{fig8}
\end{figure}

In   Fig. 8,  we show the dependence of  
$\rho_3$ and $\rho_5$  on $\varphi$, for $\lambda=0.45$, and    
$v_s=  v$,    
when   $\tan\beta=2$~(left panel) and   $\tan\beta=10$~(right panel).
From the left panel it is seen that
when $A_t=1050~\mbox{GeV}$ ($A_t=1400~\mbox{GeV}$),
the $\varphi$ dependence of 
$\rho_3$ (the upper curves) and  $\rho_5$ (the lower curves) is 
stronger  as compared to $A_t= v_s$ (the top curve with respect to the 
mid-point).
It is seen that as $\tan\beta$ 
increases, the contributions of the  CP odd components change.
For instance,  at  $\tan\beta=2$~(left panel), and $A_t=8 v_s$ 
$\rho_3$  starts from $\%98$ at $\varphi=0$ and falls to 
$\%65$ at $\varphi=\pi$ ($\rho_5$ varies from  $\%2$ to $\%37$ in this interval).
In passing to  $\tan\beta=10$  regime~(right panel) $h_5$ is seen to gain 
non-negligible $\rho_5$ composition. For instance 
at  $A_t=1400~\mbox{GeV}$,
$\rho_5$ starts with $\%50$ 
at $\varphi=0$, and  rises  to  $\%90$ at $\varphi=\pi$.
Similar to the previous observations, 
it can be observed that when  $A_t$ and $v_s$
are of comparible size, 
the strengths of the CP violation mixing is weak and
the  variations of $\rho_3$ and $\rho_5$ with $\varphi$ are quite slow.
As the figure suggests, the higher 
$A_t$
the larger the CP violating
mixings between the CP odd components. 
One notes that similar analysis can be carried out for the higher values of 
$v_s$. It can be seen that, 
as $v_s$ increases,
$\rho_5$ component of $h_3$ increases, 
and  this increase in the  $\rho_5$ component is compensated by $\rho_3$. 

\subsection{The Case of Complex $\lambda$}

In the second part of the analysis, 
we take $\lambda$ as a complex parameter,
with the  phase $\varphi_{\lambda}$.
Then, we let  all the other soft 
phases in the theory of being zero 
($\varphi_{A_t}$= $\varphi_{A_b}$=$\varphi_{A_e}$=$\varphi_{1}$= $\varphi_{2}$=0).
to explore the dependence of  $\varphi_{\lambda}$ on 
various parameters.
\begin{figure}[htb]
\vspace*{-1.6truein}
\centerline{\psfig{file=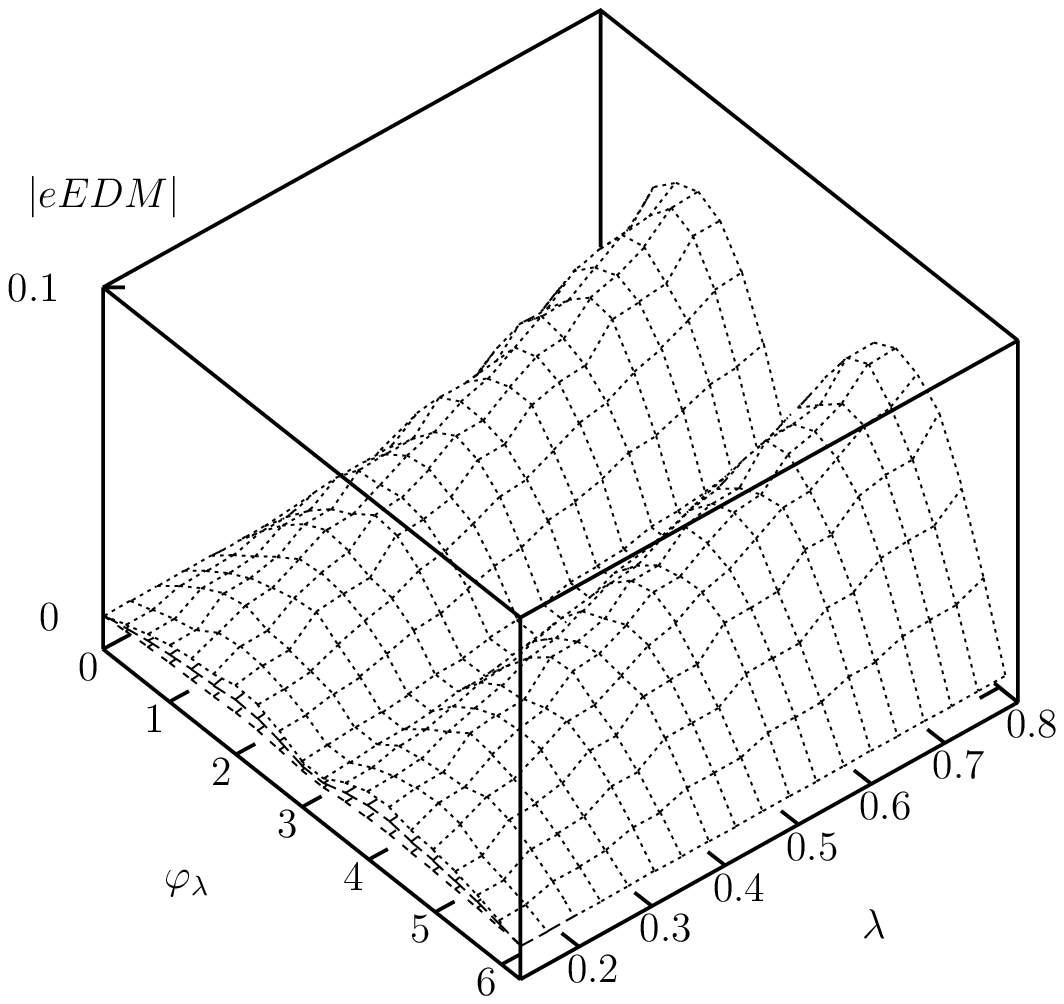,height=5.5in,width=5in }
\hspace*{-2.5truein}
\psfig{file=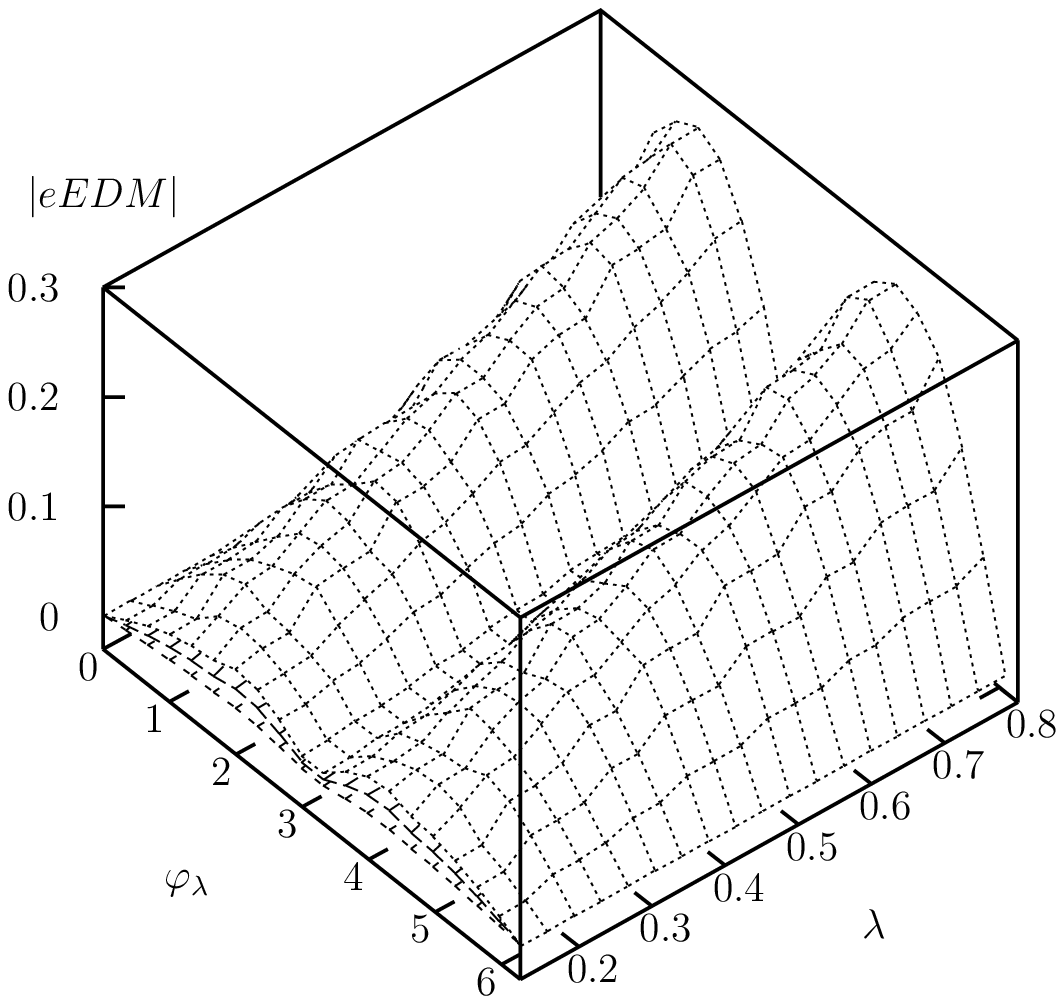,height=5.5in,width=5in }}
\vspace*{-1.7truein}
\fcaption{ The dependence of $|eEDM|$ on $\varphi_{\lambda}$ and $\lambda$ 
for selected values of the vacuum expectation value of the singlet
when $\tan\beta=2$ (left panel) and  $\tan\beta=10$ (right  panel).}
\label{fig9}
\end{figure}

In Fig. 9, we show the dependence of $|eEDM|$ on $\varphi_{\lambda}$ and $\lambda$ 
for selected values of $v_s$ 
when $\tan\beta=2$ (left panel) and  $\tan\beta=10$ (right  panel).
Similar to the former analysis, we fix $k=0.63$, and 
consider all values of  $\lambda$ in the $0.12 \simlt \lambda \simlt 0.82$,
interval, as  $\varphi_{\lambda}$ changes from 0 to $ 2 \pi$.
In both panels of the Figure, the top surface corresponds to  $v_s= 4 v$, 
whereas the bottom represents  $v_s= v$.
A comparative look at both panels of  Fig.  9 suggests that the 
present bounds of $|eEDM|$ is satisfied in all the parameter domain
at both $\tan\beta$ regimes, when  $\varphi_{\lambda}$ 
changes in its full range  and $\lambda$
from 0.12 to 0.82, provided that the gaugino masses are of ${\cal O}(\mbox{TeV})$.

In the following, we will perform a similar 
analysis that we have carried out
for the real $\lambda$ case. To understand the effects of $\varphi_{\lambda}$
on $m_{h_{5}}$, as well as $\rho_3$
and  $\rho_5$, we first start with a low value of $\lambda$, 
that is we set  $\lambda=0.12$, 
and again consider two specific values of  
$v_s$, namely  $v_s=175~\mbox{GeV}$, and  $v_s=525~\mbox{GeV}$.
We show  the dependence  of $m_{h_{5}}$  
(at  tree and at one-loop levels) on $\varphi_{\lambda}$,
in Fig. 10 (when $v_s= v$), and Fig. 11 ($v_s= 3 v$) 
at   $\tan\beta=2$~(left panels) and   $\tan\beta=10$~(right panels).
\begin{figure}[htb]
\vspace*{-2.1truein}
\centerline{\psfig{file=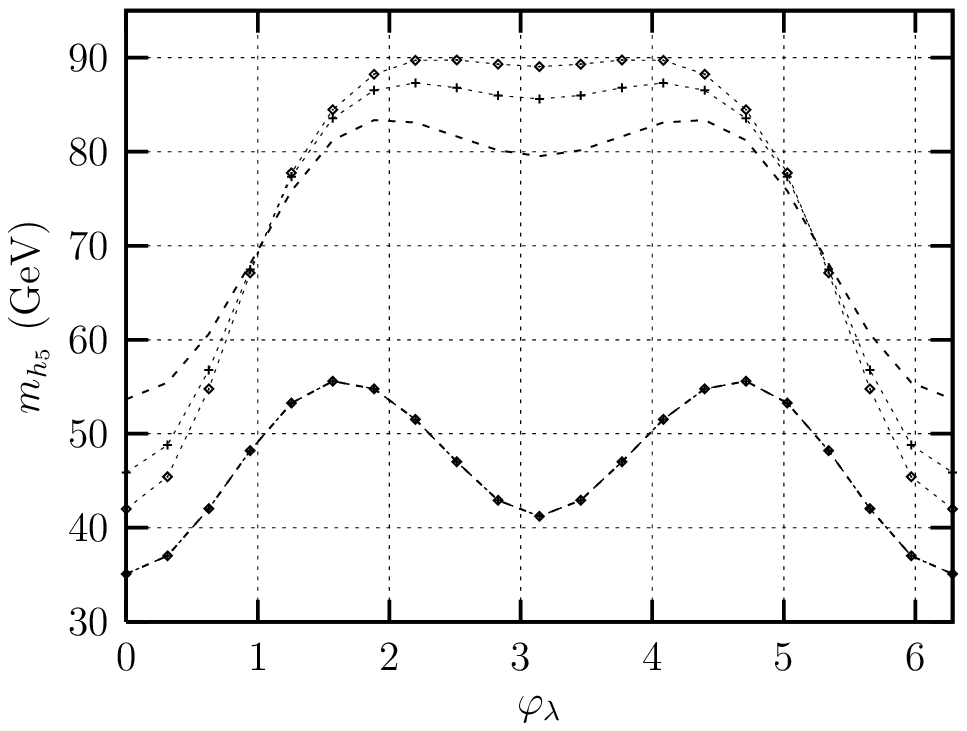,height=6.0in,width=4.8in }
\hspace*{-2.5truein}
\psfig{file=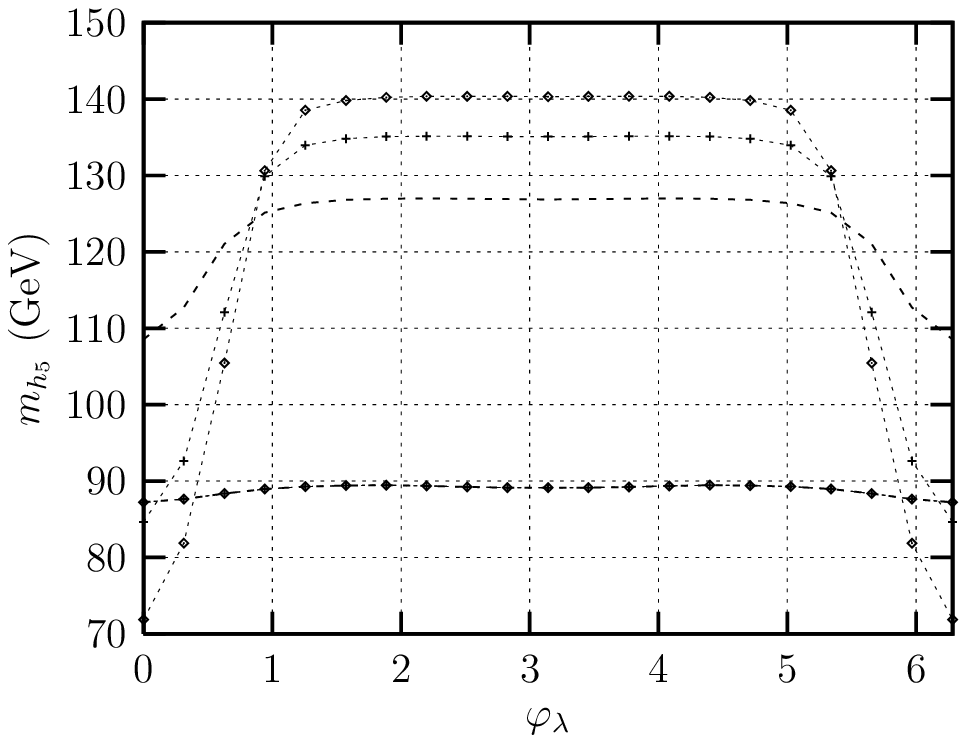,height=6.0in,width=4.8in }}
\vspace*{-2.3truein}
\fcaption{ The dependence of  
$m_{h_{5}}$
on $\varphi_{\lambda}$, for selected values of $A_t$,
at $\tan\beta=2$~(left panel) and $\tan\beta=10$~(right panel), 
when  $v_s=175~\mbox{GeV}$, v and  $\lambda$= 0.12.
Here, the bottom, middle and the top curves (with respect to the mid-point)
are for $A_t=v_s=175~\mbox{GeV}$, $A_t=1050~\mbox{GeV}$, and   
$A_t=1400~\mbox{GeV}$, respectively, whereas the  lowest curve is for the tree-level.}
\label{fig10}
\end{figure}

As the left panel of Fig. 10 indicates   $m_{h_{5}}$
is quite sensitive to the variations of  $\varphi_{\lambda}$ when
$v_s=v$, and  $\tan\beta=2$.
For  $\tan\beta=10$ regime, it is seen that the variation 
of  $\varphi_{\lambda}$ with  $m_{h_{5}}$  is slower
at both tree and one-loop levels,
particularly in the $1 \simlt \varphi_{\lambda} \simlt 5$ 
interval. 
\begin{figure}[htb]
\vspace*{-2.1truein}
\centerline{\psfig{file=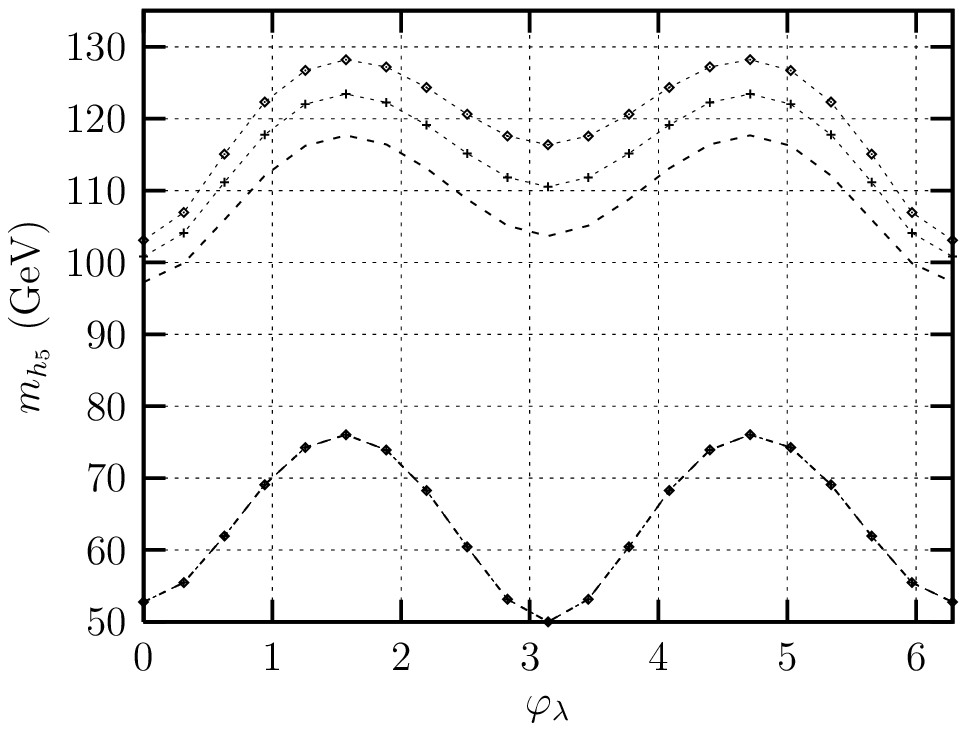,height=6.0in,width=4.8in }
\hspace*{-2.5truein}
\psfig{file=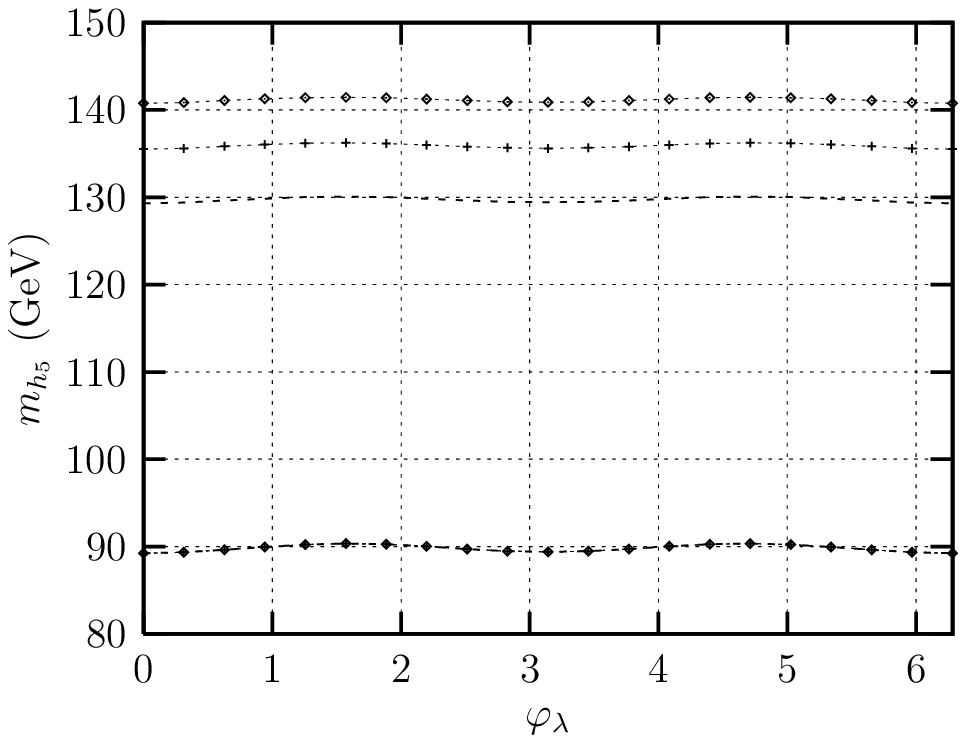,height=6.0in,width=4.8in }}
\vspace*{-2.3truein}
\fcaption{ The dependence of  
$m_{h_{5}}$
on $\varphi_{\lambda}$, for 
selected values of $A_t$, at 
$\tan\beta=2$~(left panel) and $\tan\beta=10$~(right panel), 
when  $v_s=525~\mbox{GeV}$ and  $\lambda$= 0.12.  
Here, the  bottom, the middle and the top curves,      
are for $A_t=v_s=525~\mbox{GeV}$,
$A_t=1050~\mbox{GeV}$, and   $A_t=1400~\mbox{GeV}$, respectively, 
whereas the  lowest curve is for the tree level.} 
\label{fig11}
\end{figure}

In passing to  $v_s=3 v$
case (see Fig. 11),  although $m_{h_{5}}$  remains quite flat 
for $\tan\beta=10$~(right panel),
one obtains a relatively fast variation for  $\tan\beta=2$.
It is seen that from the left panel of Fig. 11 that
the peak value 
of $m_{h_{5}}$ at the one-loop level, as well as the tree level,
is shifted through the maximal CP violation point at $\tan\beta=2$.
A comparative analysis of Figs. 10 and 11 suggest that  $m_{h_{5}}$  
is quite sensitive to the 
variations in $\varphi_{\lambda}$ for 
$\lambda=0.12$, when  $v_s=v$ and  $v_s= 3 v$ at    
$\tan\beta=2$~(left panels) regime. 
Such kind of sensitivity weakens at  
$\tan\beta=10$~(right panels)
for  $v_s \simgt  2 v$. 

We show the dependence of  CP-odd components 
($\rho_3$ and $\rho_5$), of $h_5$
on $\varphi_{\lambda}$, for $\lambda=0.12$, and    
$v_s=  v$ in Fig. 12,  and   $v_s= 3 v$ in Fig. 13, 
respectively,
when   $\tan\beta=2$~(left panels) and   $\tan\beta=10$~(right panels).
\begin{figure}[htb]
\vspace*{-2.1truein}
\centerline{\psfig{file=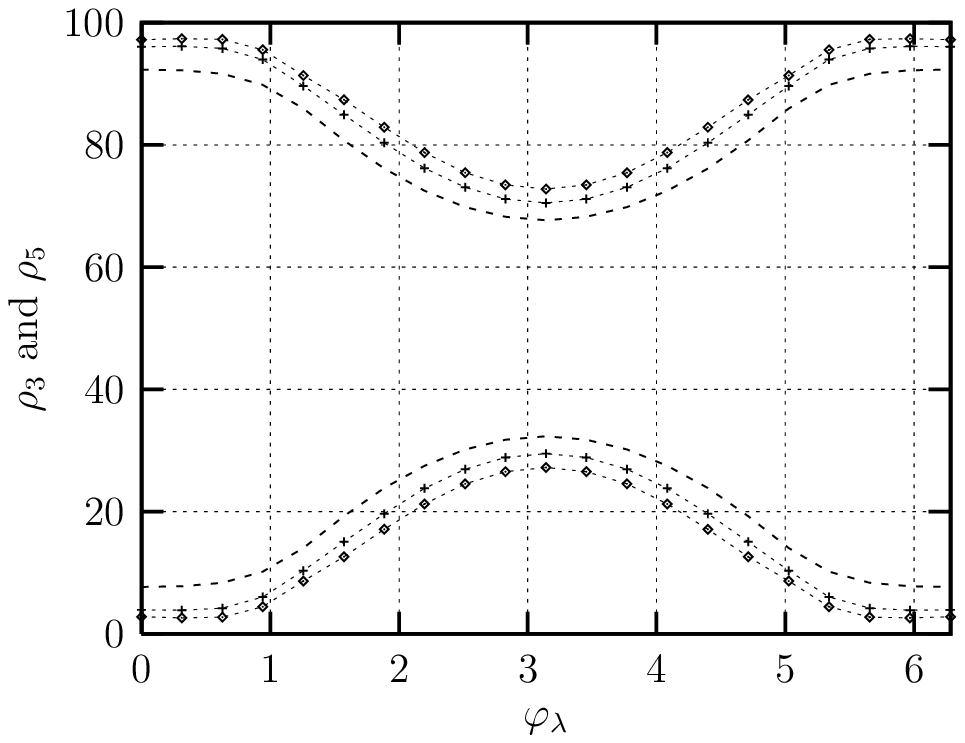,height=6.0in,width=4.8in }
\hspace*{-2.5truein}
\psfig{file=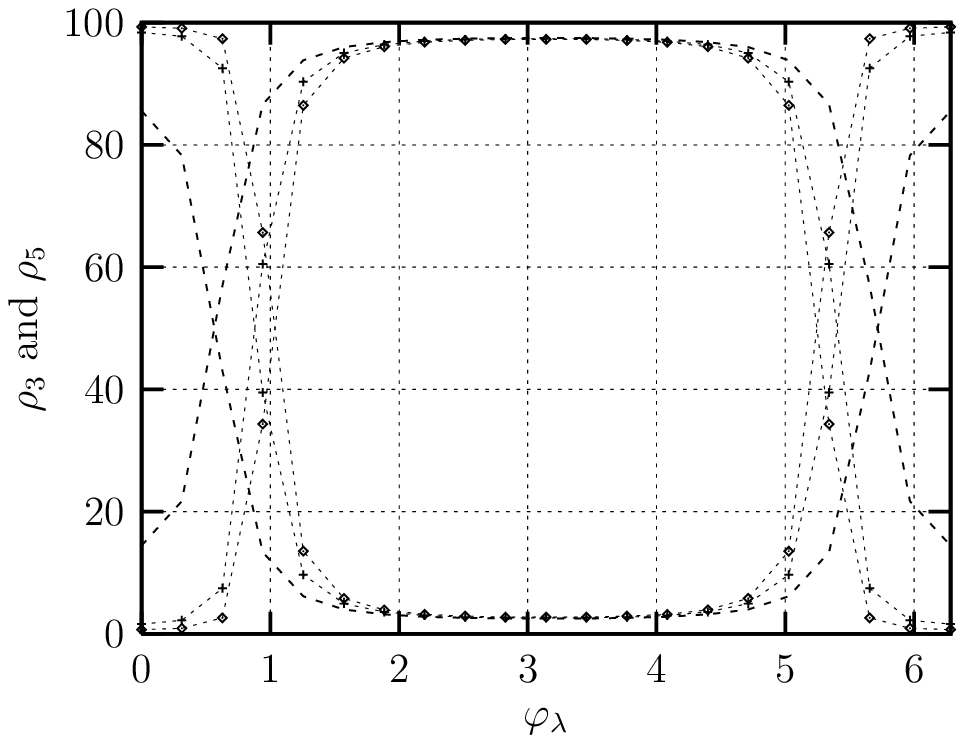,height=6.0in,width=4.8in }}
\vspace*{-2.3truein}
\fcaption{ The dependence of   
$\rho_3$ (the upper curves) and $\rho_5$ (the lower curves)
on $\varphi_{\lambda}$ for  $v_s=175~\mbox{GeV}$, 
and $\lambda$= 0.12,
when $\tan\beta=2$ (left panel), and $\tan\beta=10$ (right panel).
Here,   $A_t=v_s=175~\mbox{GeV}$, $A_t=1050~\mbox{GeV}$,  $A_t=1400~\mbox{GeV}$, 
from bottom to top for $\rho_3$, and from top  to bottom for  $\rho_5$, with
respect to the mid-point.} 
\label{fig12}
\end{figure}
As can be observed from the left panel of Fig. 12,  
when  $\tan\beta=2$, $\rho_3$  component of $h_5$ on the average 
ranges from $\%98$ to $\%90$,
as $A_t$ changes from   $1400~\mbox{GeV}$ to    $175~\mbox{GeV}$ 
(from top to bottom for the upper lines) at
$\varphi_{\lambda}=0$. In passing to  $\tan\beta=10$ regime, it is seen that    
$h_5$ has nearly $\%100$  $\rho_3$ component at  $\varphi_{\lambda}=0$, as $A_t$ changes from
$1400~\mbox{GeV}$ to  $1050~\mbox{GeV}$ 
(the first and the second curve from the top of  the upper lines representing $\rho_3$). 
Moreover, $h_5$ is seen to gain non-negligible  $\rho_5$ component 
in such a way that  $\rho_3$ and  $\rho_5$ components are strongly mixed up
in this regime~(right panel). 
\begin{figure}[htb]
\vspace*{-2.1truein}
\centerline{\psfig{file=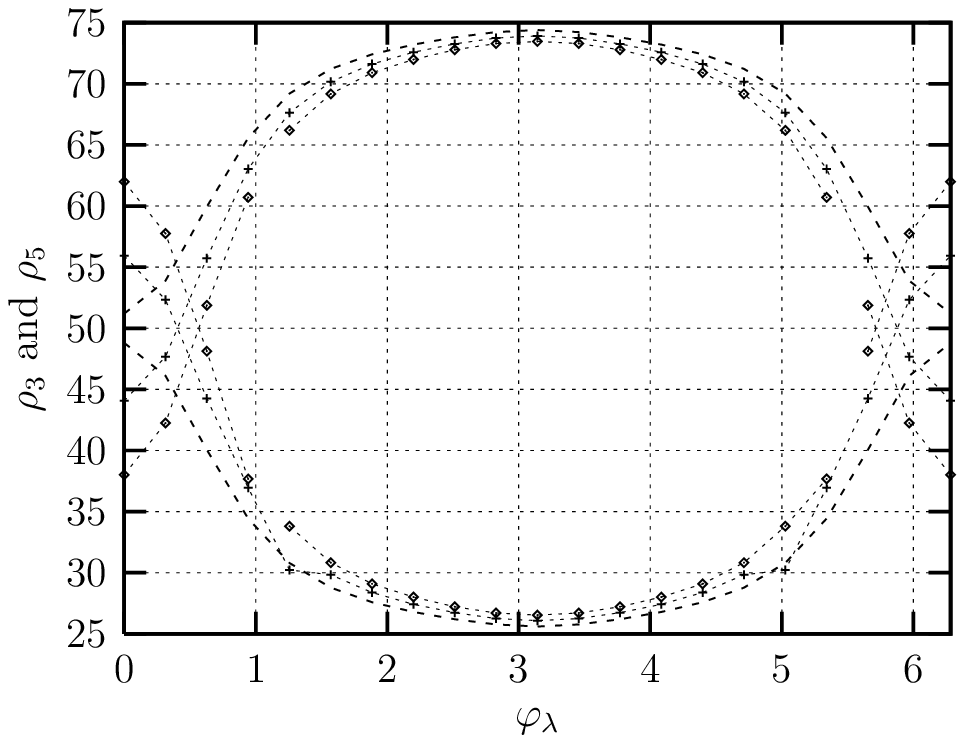,height=6.0in,width=4.8in }
\hspace*{-2.5truein}
\psfig{file=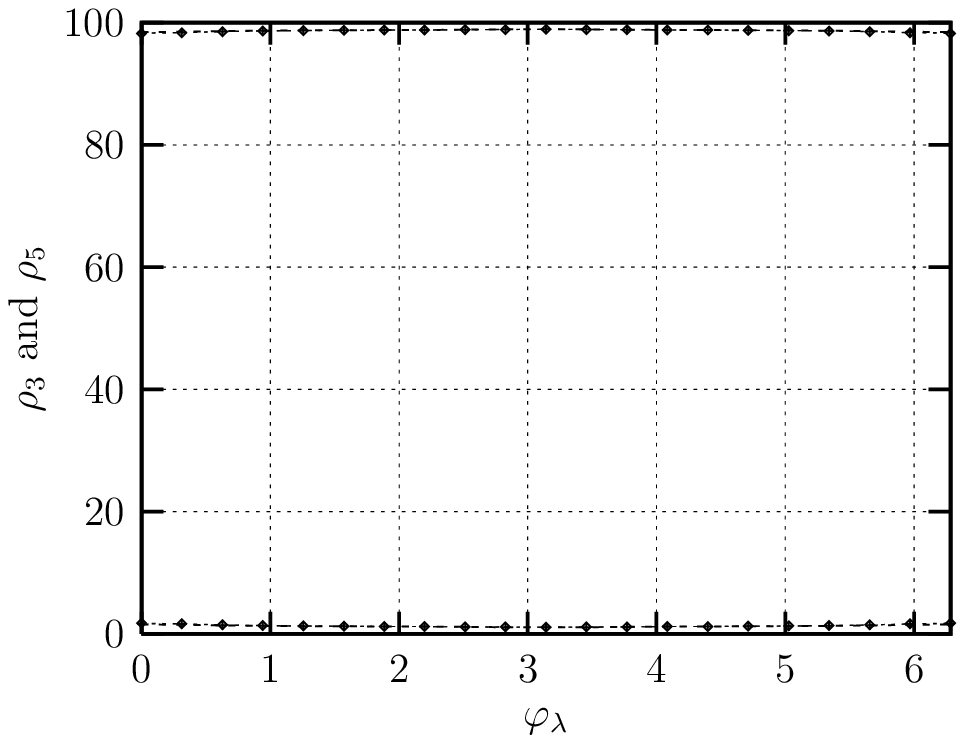,height=6.0in,width=4.8in }}
\vspace*{-2.3truein}
\fcaption{ The dependence of  
$\rho_3$ (the lower curves) and $\rho_5$ (the upper curves)
on $\varphi_{\lambda}$ for  $v_s=525~\mbox{GeV}$, 
and $\lambda$= 0.12,
when $\tan\beta=2$ (left panel), and $\tan\beta=10$ (right panel).
Here,   $A_t=v_s=525~\mbox{GeV}$, $A_t=1050~\mbox{GeV}$,  $A_t=1400~\mbox{GeV}$, 
from bottom to top for $\rho_3$, and from top to bottom for  $\rho_5$, with
respect to the mid-point.} 
\label{fig13}
\end{figure}

On the other, considering the higher values of $v_s$ (see Fig. 13), 
one notes that 
the $\rho_5$ component of $h_5$ increases,
and this increase in the  $\rho_5$ is compensated by $\rho_3$
when  $\tan\beta=2$~(left panel).
For  $\tan\beta=10$~(right panel) regime,
it can be observed that   $\rho_5$ 
component of $h_5$ has a slow variation in the vicinity of  $\%100$ line,
whereas $\rho_3$ remains around  $\%0$ line, for each value of $A_t$,  
changing from  $525~\mbox{GeV}$ to  $A_t=1400~\mbox{GeV}$.

In the last  part of our analysis, we fix  $\lambda=0.45$, and carry out 
the  similar analysis for    $v_s=v$ and  $v_s=3 v$. 
We show  the dependence  of  $m_{h_{5}}$ 
(at  tree and at one-loop levels) on $\varphi_{\lambda}$ for    $v_s=v$   
in Fig. 14, and   $v_s=3  v$  in Fig. 15,  when   $\tan\beta=2$~(left
panels) and   $\tan\beta=10$~(right panels).
In Figs. 14 and 15, $A_t=1400~\mbox{GeV}$~(the top curve with respect to the mid-point),
$A_t=1050~\mbox{GeV}$~(the second curve below the top curve),
whereas the third curve below the top  curve
is for $A_t=v_s=175~\mbox{GeV}$ in Fig. 14, and  $A_t=v_s=525~\mbox{GeV}$
in Fig. 15. In both Figures, the  lowest curve 
represents $m_{h_{5}}$ at the tree-level.
\begin{figure}[htb]
\vspace*{-2.1truein}
\centerline{\psfig{file=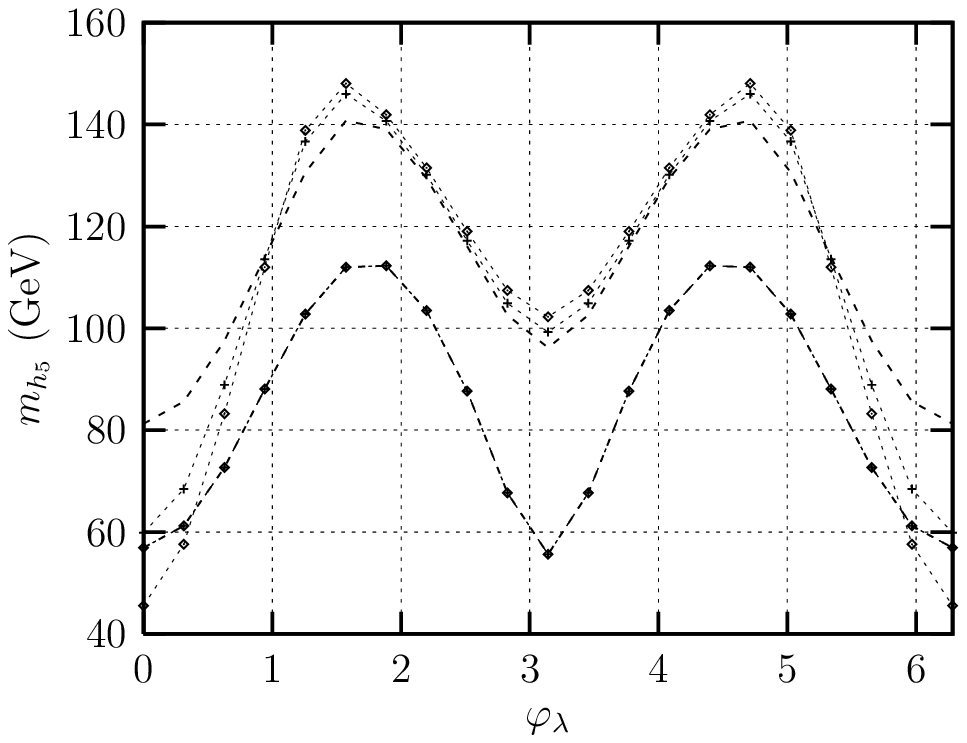,height=6.0in,width=4.8in }
\hspace*{-2.5truein}
\psfig{file=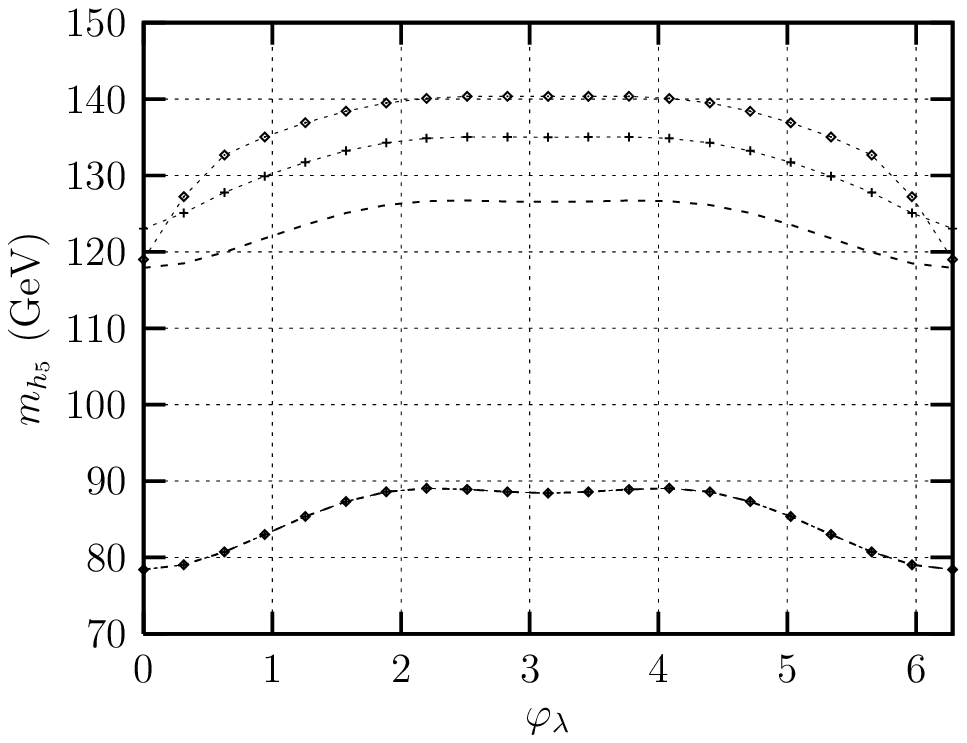,height=6.0in,width=4.8in }}
\vspace*{-2.3truein}
\fcaption{ The dependence of  
$m_{h_{5}}$
on $\varphi_{\lambda}$ for  $v_s$= v, 
and $\lambda$= 0.45, for selected values of $A_t$,
when $\tan\beta=2$ (left panel), and $\tan\beta=10$ (right panel).
Here, the  bottom, the middle and the top curves,      
with respect to the mid-point, are for $A_t=v_s=175~\mbox{GeV}$,
$A_t=1050~\mbox{GeV}$, and   $A_t=1400~\mbox{GeV}$ values, 
whereas the  lowest curve is for the tree level.} 
\label{fig14}
\end{figure}
\begin{figure}[htb]
\vspace*{-2.1truein}
\centerline{\psfig{file=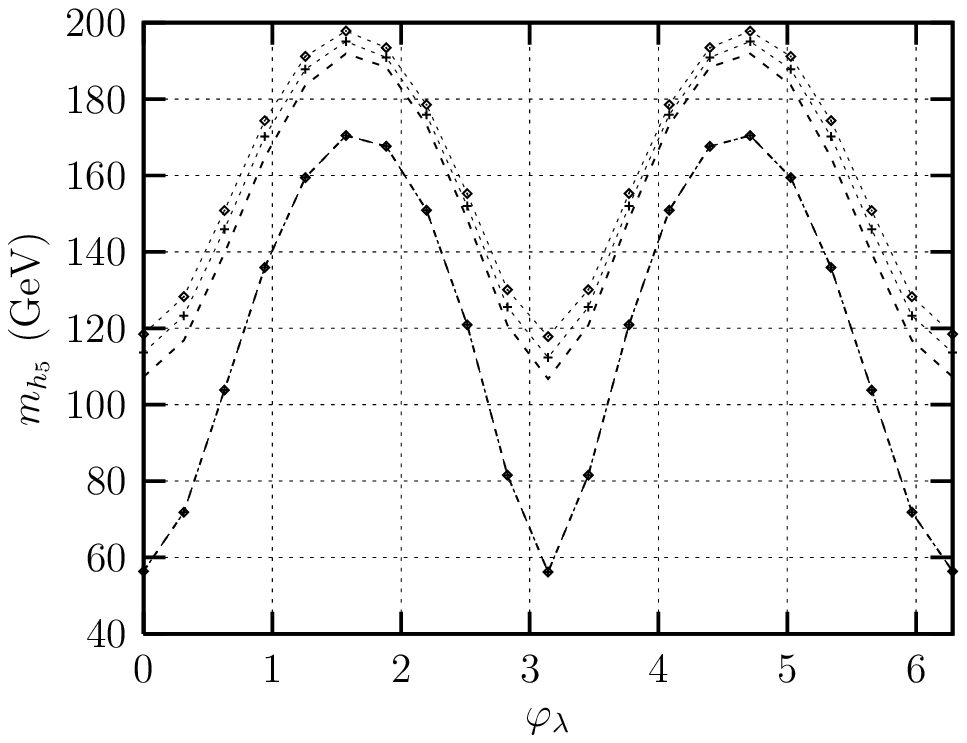,height=6.0in,width=4.8in }
\hspace*{-2.5truein}
\psfig{file=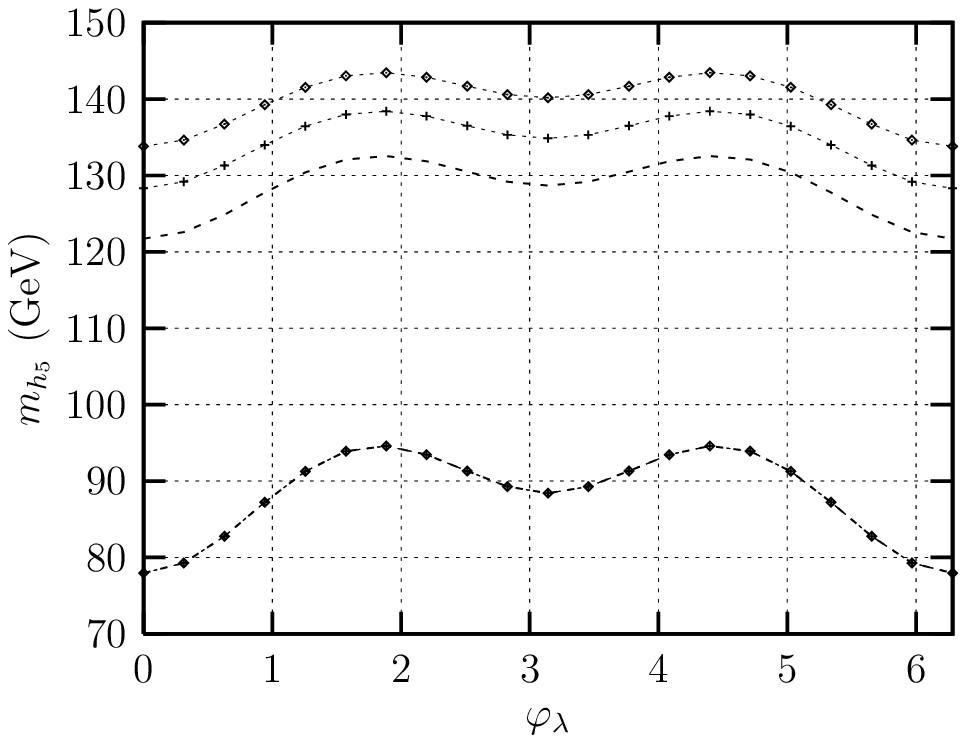,height=6.0in,width=4.8in }}
\vspace*{-2.3truein}
\fcaption{ The dependence of  $m_{h_{5}}$ on $\varphi_{\lambda}$ for  $v_s$=3 v, 
and $\lambda$= 0.45, for selected values of $A_t$,
when $\tan\beta=2$ (left panel), and $\tan\beta=10$ (right panel).
Here, the  bottom, the middle and the top curves,      
with respect to the mid-point,
are for $A_t=v_s=525~\mbox{GeV}$,
$A_t=1050~\mbox{GeV}$, and   $A_t=1400~\mbox{GeV}$ values, 
whereas the  lowest curve is for the tree level.} 
\label{fig15}
\end{figure}

One notes from the left panels of Figs. 14 and 15 that
as $\lambda$ increases ($\lambda=0.45$), 
one obtains a quite fast variation of $m_{h_{5}}$ with $\varphi_{\lambda}$
both at the tree and loop levels, particularly in the low-$\tan\beta$
regime. 
The increase in $\lambda$, as well as $v_s$
affects the spectrum of ${h_{5}}$ in the sense that  
the shifting of the  peak value of $m_{h_{5}}$ (at the tree and the loop levels)
towards the maximal CP violation point is  sharper
as compared to the $\lambda=0.12$ case (see Fig. 11). 
Although the sensitivity much more weakens in the high $\tan\beta$ regime,
such kind of shifting effect 
of $m_{h_{5}}$ can also be observed from the right panel of 
Fig. 15.

Finally, in Fig. 16, we show the dependence of  CP-odd components 
($\rho_3$ and $\rho_5$) of $h_3$
on $\varphi_{\lambda}$, for $\lambda=0.45$,     
and $v_s=  v$  at   $\tan\beta=2$~(left panel) and   $\tan\beta=10$~(right panel).
As can be observed from  Fig. 16, 
the increase in $\lambda$ affects the 
CP odd components  in such a way that $\rho_3$
and  $\rho_5$  mix significantly in the $\tan\beta=2$ regime.
In passing to $\tan\beta=10$ regime, it is seen that such mixings weaken,
however  $h_5$ is seen to gain non-negligible $\rho_5$ component
as  $A_t$ increases. On the other hand, 
as $v_s$  increases, it can be  seen that 
$\rho_3$ and   $\rho_5$ have similar $\varphi_{\lambda}$
dependencies for all values of $A_t$, which essentially follows from the 
dominance of the $ A_t \lambda v_s$ term.
\begin{figure}[htb]
\vspace*{-2.1truein}
\centerline{\psfig{file=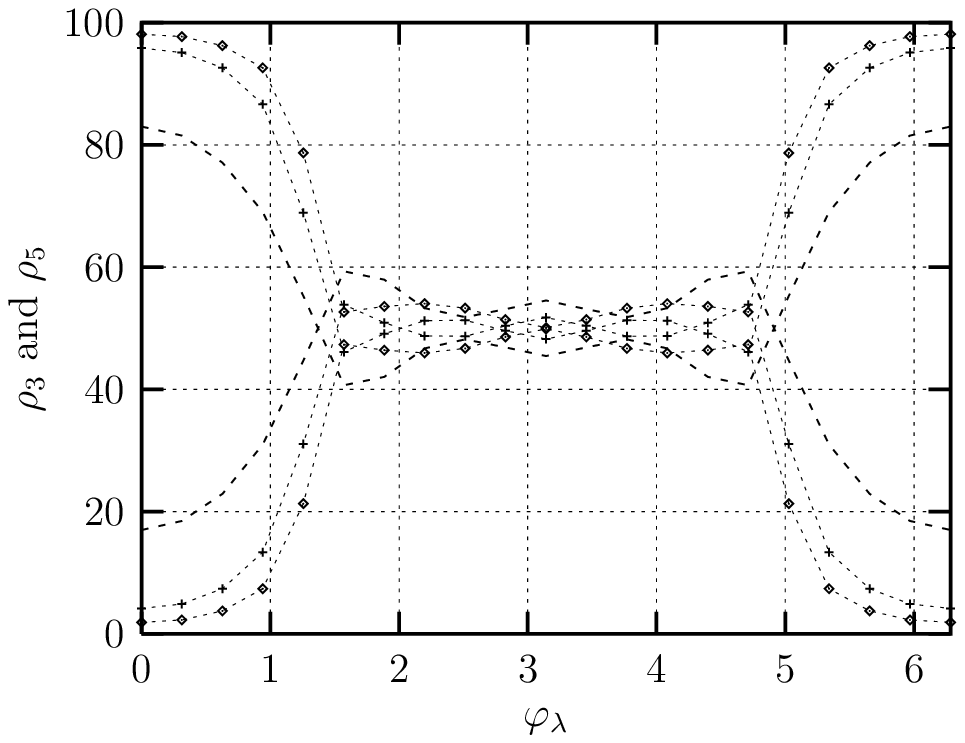,height=6.0in,width=4.8in }
\hspace*{-2.5truein}
\psfig{file=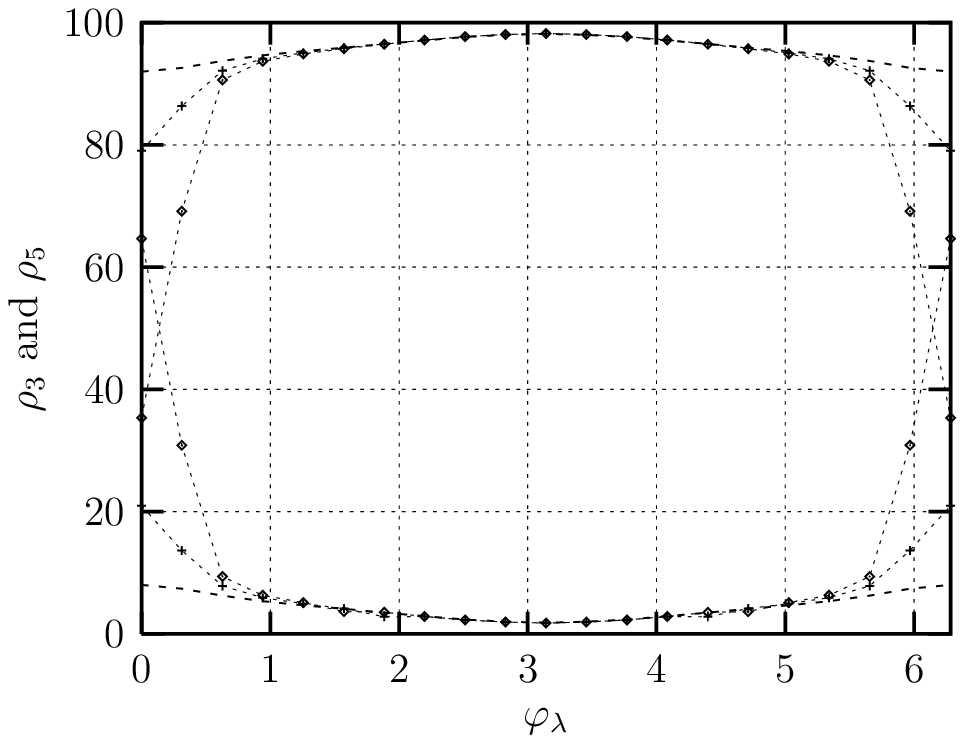,height=6.0in,width=4.8in }}
\vspace*{-2.3truein}
\fcaption{The dependence of 
$\rho_3$  and $\rho_5$ on $\varphi_{\lambda}$ for  $v_s$= v, 
and $\lambda$= 0.45, for selected values of $A_t$,
when $\tan\beta=2$~(left panel), and $\tan\beta=10$~(right panel).
The upper (lower) curves are for $\rho_3$($\rho_5$) in the left panel,
whereas they are for  $\rho_5$( $\rho_3$)  in the right  panel.
Here,   $A_t=v_s=175~\mbox{GeV}$, $A_t=1050~\mbox{GeV}$,  $A_t=1400~\mbox{GeV}$, 
from top to bottom for $\rho_3$ ($\rho_5$), with respect to the mid-point.} 
\label{fig16}
\end{figure}

Before concluding, we would like to close this section 
with a brief discussion of the 
measurement of CP violating effects at  colliders:
Recently, the authors of Ref. [44]
have presented a general formalism,
the  resonant CP violation, 
for analyzing the 
CP violating phenomena in the MSSM at high-energy colliders.
Their formalism, which is developed from Ref.~[45],
can be applied to models with an extended CP violating Higgs sector,
including  the MSSM with radiative Higgs-sector CP violation.
Resonant CP violation has not been studied in the NMSSM yet.
The measurement of  resonant CP violation 
effects
can provide a 
useful tool for the 
determination 
of the CP odd compositions of the Higgs bosons in the 
NMSSM. 

\section{Conclusion}

We study the explicit CP violation of the Higgs sector in the 
next--to--minimal supersymmetric model with a gauge singlet Higgs field.
Our general discussion followed by the numerical estimates for various 
parameter planes  show that:

($i$)The eEDM  
lies around the present experimental upper limits, provided that the gaugino masses are of ${\cal O}(\mbox{TeV})$.

($ii$)When  $\lambda$ is  real, and all the other phases in the theory are 
complex, 
$m_{h_{5}}$ 
is quite sensitive to $\varphi$,
particularly at small $\lambda$ and $v_s$ ($\lambda=0.12$, and $v_s=v$).
However as $v_s$ and  $\tan\beta$  increases 
(for instance, $\lambda=0.12$, $v_s=3
v$ and $\tan\beta=10$), the radiative corrections which are sensitive to 
variations in $\varphi$ 
are suppressed. Indeed,  $m_{h_{5}}$ nearly remains constant for  
$\tan\beta=10$ at $v_s=3 v$. 
Considering the   $\rho_3$ and $\rho_5$  components of $h_5 $
at small $\lambda$ and $v_s$ ($\lambda=0.12$, and $v_s=v$),
as $A_t$
increases, the $\rho_3$ component of $h_5 $ decreases,
whereas its $\rho_5$ component increases.
However,  this increase in $\rho_{5}$ can not be larger than  $\%20$, at
$\tan\beta=2$, even for 
$A_t=1400~\mbox{GeV}$.
In passing to   $\tan\beta=10$ regime, it is seen that 
the $\rho_3$ component of $h_5$ tend to decrease more rapidly,
whereas this decrease  is compensated by the increase  in  $\rho_5$,
in the sense that the two components  mix  
around  $\varphi=\pi$, when $\lambda=0.12$. On the other hand, 
for higher values of $v_s$ ($v_s=3 v$),
it is seen that the variation of the CP odd components is quite slow, as
compared to the former case. 

($iii$) When $\lambda$ is complex,
and all the other CP violating parameters in the theory are chosen to be real,
$m_{h_{5}}$ 
grows significantly as compared to the
former case. It is observed that $\varphi_{\lambda}$
strongly  affect the the tree and one-loop Higgs masses. 
Clearly,  $\varphi_{\lambda}$ is not only a physical phase
which forms the  source of CP violation at the tree level, but it also 
affects the one-loop radiative corrections  via  $\varphi_{\lambda t}$.
Therefore, even if all the other phases are set equal to zero.
the effects of $\varphi_{\lambda}$ can be seen on the  mass
and the CP components of $h_5$. 
The CP violating effects are particularly enhanced, 
as $v_s$ increases.

($iv$)Being  the most economic extension of the MSSM,   
the next-to-minimal supersymmetric model not only leads us to a 
wealth of CP violation opportunities, but
it also offers a rich phenomenology  for future colliders.

\end{document}